\documentclass[a4paper,11pt]{article}
\usepackage{amsmath}
\usepackage{amssymb}
\usepackage[pdftex]{graphicx}
\usepackage{epstopdf}
\usepackage{longtable}
\usepackage{subfigure}
\usepackage{natbib}
\usepackage[pdftex, pdftitle={On the Running Time of the Shortest Programs},
pdfauthor={Norbert Batfai},
pdfkeywords={Kolmogorov complexity, Busy Beaver, Placid Platypus},
pdfstartview=FitB]{hyperref}
\usepackage{amssymb,amsmath,amsthm,amsfonts}

\newtheorem{theorem}{Theorem}
\newtheorem{definition}{Definition}
\newtheorem{notation}{Notation}
\newtheorem{remark}{Remark}
\newtheorem{example}{Example}

\newcommand{\keywords}[1]{\par\addvspace\baselineskip\noindent\textbf{Keywords: }\textit{#1}.}

\author{Norbert B\'atfai
\\University of Debrecen
\\Department of Information Technology
\\\texttt{batfai.norbert@inf.unideb.hu}}
\title{On the Running Time of the Shortest Programs}
\begin{document}
\maketitle

\begin{abstract}
The Kolmogorov complexity of the word $w$ is equal to the length of the shortest concatenation of program $Z$ and its input $x$ with which the word $w$ is computed by the universal turing machine $U$. The question introduced in this paper is the following: How long do the shortest programs run for?
\keywords{Kolmogorov complexity, Busy Beaver, Placid Platypus}
\end{abstract}

\tableofcontents
%\listoffigures
%\listoftables

\section{Introduction}

\subsection{Combining Kolmogorov Complexity and Running Time}

The Kolmogorov complexity of the word $w$ is 
$K(w)=min\{|Zx| \big| U(Zx)=w\}$. This famous concept of complexity is quite near to our intuition if compared to other measures of complexity (for example Shannon entropy and computational complexity) because it is equal to the length of the shortest concatenation of program $Z$ and its input $x$ with which the universal turing machine $U$ produces the word $w$. This definition may be made more precise, for example by using a separator in the concatenation or assuming that $Z$ and $x$ are self delimiting words. A detailed discussion of this theme can be found in the book \citep{LiVit} or a shorter introduction can be found in the Hungarian-language book \citep{Ronyai}.

The main question introduced in this paper is the following: How long does it take to run the shortest programs?  In the general case we define the time complexity of word $w$ as follows 
$K_t(w)=min\{t_U(Zx) | U(Zx)=w\}$ where $t_U(Zx)$ denotes the number of steps of the universal Turing machine $U$ started with input $Zx$. Now restrict this general definition for the shortest programs. Define the Kolmogorov time complexity of word $w$ by 
$K_T(w)=min\{t_U(Zx) | U(Zx)=w, |Zx|=K(w)\}$.

The basic properties of these notions are extremly simple, 
$|w| \le K_t(w) \le K_T(w)$ because we have to print $w$ letter by letter and $K_T(w)$ is defined on a subset of the same set. It is clear from the operating of the universal machine $U$ that there exist constants $c_{init}$ and $c_{run}$ that
\[t_U(Zx) \le 
\underbrace{c_{init}|Zx|}_\text{initializing of U} + 
\underbrace{c_{run} t_Z(|x|)}_\text{U simulates Z}\]
where $t_Z(n)$ is the usual time complexity function. By using this relation it is easy to see that there exists a constant $c$ that $K_t(w) \le c|w|$ because 
 $K_t(w) \le c_{init}|Z_{trivial}w|$, where $Z_{trivial}$ is an immediately halting machine that is accepting anything, so $U(Z_{trivial}w)=w$ and for every $n$ holds $t_{Z_{trivial}}(n)=0$.

There may exist in principle extreme incompressible word w such that 
$K(w) = |Z_{trivial}w|$.  Among these words we may have trivially restricted ourselves to the case where the shortest program is identical to the trivial program.

\begin{theorem}
$K(w) = |w|+|Z_{trivial}| \Rightarrow  K_T(w)=O(1)K(w).$
\end{theorem}

\begin{proof}
This holds trivially because $K_T(w) \le c_{init}|Z_{trivial}||w|$
\end{proof}

Can we make similar statements for the cases of $|w| \le K(w)$ incompressible and  $|w| > K(w)$ compressible words? If we can do then we would not violate the undecidability of the halting problem because the used Kolmogorov complexity function $K$ is not recursive (and on the other hand the shortest programs are always halting programs). We have conjectured that the shortest program cannot run on indefinitely in time.

In the following we shall restrict ourselves to the case of Busy Beaver words, because the general case seems to be quite difficult. These words are very well compressible, that is $|w| >> K(w)$.

We now give a brief outline of the content of this paper. In the next section, we introduce an usual Turing machine model for Busy Beaver problem. In section 3, we consider the running time of some Busy Beaver machines. Finally, in appendixes we show results of our searching programs.

\section{Busy Beaver machines work like the Shortest Programs}

The Busy Beaver problem, introduced by Tibor Rad\'o more than 40 years ago, is to find the $n$-state, binary tape Turing machine that is starting with the empty word and has written the utmost number of ones on the tape. 

\subsection{Notations and programs}

In this paper and our Busy Beaver computing programs we will use
the following variant of definition of Turing machine. 
\begin{definition}[Busy Beaver words]\label{TuringMachine}
A Turing machine
is a quadruple $T=(Q_T, 0, \{0,1\}, f_T)$ 
where 
$Q_T \subset \mathbb{N}$ is the state alphabet, 
$0 \in Q_T$ is the initial state, 
the tape alphabet is binary 
and finally 
$f_T$ is the partial transition function 
$f_T:Q_T\times\{0,1\} \rightarrow Q_T\times\{0,1\}\times\{\leftarrow,\uparrow,\rightarrow\}$.
If $f$ is undefined for some pair $(q, c)$ then the machine $T$ is
going to halt.
The word $w$ computed by machine $T$, denoted $T(\lambda)=w$, is the concatenation of tape symbols from the leftmost symbol 1 to the rightmost symbol 1 provided that $T$ has halted.
We will say that $T \in BB_n$ provided that $Card(Q_T) \le n$. 
The Busy Beaver function, introduced by Rad\'o, is defined as 
$BB(n) = max\{\text{nr\_ones}(w) \big| T(\lambda)=w, T \in BB_n\}$, here and in the following $\text{nr\_ones}(w)$ denotes the function $\sum_{i=1}^k{w_i}, w=w_1\dots w_k$. The machines and words which belong to the maximum will be called Busy Beaver machines and Busy Beaver words. 
\end{definition}

\begin{notation}
Elements of the transition function are often referred to as transition rules. 
For example, let $T=(\{0, 1, 2, 3\}, 0, \{0,1\}, 
\{
(0,0, 0,1,\leftarrow),
(0,1, 0,0,\rightarrow), 
(1,0, 0,0,\leftarrow), 
(1,1, 0,1,\leftarrow), 
(2,0, 0,1,\rightarrow),
(3,0, 1,1,\leftarrow)
\})$.
The program of this machine $T$ consist of six rules 
\[
\begin{aligned}
(0,0) &\rightarrow (0,1,\leftarrow) \\
(0,1) &\rightarrow (0,0,\rightarrow) \\
(1,0) &\rightarrow (0,0,\leftarrow) \\
(1,1) &\rightarrow (0,1,\leftarrow) \\
(2,0) &\rightarrow (0,1,\rightarrow) \\
(3,0) &\rightarrow (1,1,\leftarrow)
\end{aligned}
\]
But in most cases it is easiest to use the state transition diagram when we are interpreting the operation of a machine. A such diagram of this machine $T$ is shown in Figure \ref{bb4m8}. 
In this diagram, states are nodes and edges are labelled with 
$(read, write, move)$-shaped triples. 

The word $w=T(\lambda)$ computed by this machine $T$ is $1010111111$.
The computing of this word is shown in the following rows, where the \texttt{[q>} denotes the position of the head on the tape and the state $q$ of the machine $T$.

\begin{verbatim}
Step 0     00000000[0>00000     0000000000000
Step 1     0000000[1>010000     0000000010000
Step 2     000000[3>0010000     0000000010000
Step 3     00000[2>01010000     0000001010000
Step 4     000001[0>1010000     0000011010000
Step 5     0000010[2>010000     0000010010000
Step 6     00000101[0>10000     0000010110000
Step 7     000001010[2>0000     0000010100000
Step 8     0000010101[0>000     0000010101000
Step 9     000001010[1>1100     0000010101100
Step 10    00000101[0>01100     0000010101100
Step 11    0000010[1>111100     0000010111100
Step 12    000001[0>0111100     0000010111100
Step 13    00000[1>11111100     0000011111100
Step 14    0000[0>011111100     0000011111100
Step 15    000[1>0111111100     0000111111100
Step 16    00[3>00111111100     0000111111100
Step 17    0[2>010111111100     0010111111100
Step 18    01[0>10111111100     0110111111100
Step 19    010[2>0111111100     0100111111100
Step 20    0101[0>111111100     0101111111100
Step 21    01010[2>11111100     0101011111100
\end{verbatim}

We will denote a Turing machine by $(n, f_1, t_1, \dots ,f_n, t_n)$ where $f_i \rightarrow t_i$ is a transition rule, the $f_i$ is the from part of the rule, $t_i$ is the to part of the rule. These parts are shown in code snipets in Appendix \ref{codesnips}. For example, the name of the machine $T$ is $(6, 0, 9, 1, 14, 2, 18, 3, 3, 4, 5, 6, 15)$.

\end{notation}

\begin{remark}
For a given number of states $n$ and given number of rules $k$, the number of machines consisting of $k$ rules is $\binom{2n}{k}(6n)^k$, so the total number of machines with n states is  
\[
\sum_{k=1}^{2n}\binom{2n}{k}(6n)^k
\]
by using Newton's binomial theorem, we obtain $(6n+1)^{2n}-1$. 

If $n=4$ then it equals $152.587.890.624$. Thus, assuming the lower and upper bound of the time needed to simulate a $BB_4$ machine are $33$ $\mu s$ and 
$76$ $\mu s$\footnote{These values are based on our brute force simulation programs which are enumerating and simulating all $BB_4$ machines. The times 2 system call was used to measure the elapsed time for the simulation of a $BB_4$ machine.}. It means that a ''brute force'' program on a PC with $6100$ BogoMips will been computing for number of days from $58$ to $135$ if we would like to simulate all $BB_4$ machines. The same values for the case $n=5$ are $13775$ years and $57179$ years provided that the minimal and maximal simulating times of a $BB_5$ machine are  $0.53$ $ms$ and $2.2$ $ms$\footnote{This two estimations were made with naive low value of maximum number of steps. It was equals to only 32000, but then, for example the number of steps of the winner candidate (Marxen-Buntrock) machine is more than 40 million steps.}.

It appears to be very difficult to give estimation and so to write programs which are working well in the case $n \ge 6$ because we have no any theoretical limit values (for example maximal number of used tape cells or maximal number of steps) that we may use in some simulation program.
\end{remark}

\begin{remark}
We should remark that the definition of Turing machine used by us may changes the Busy Beaver function. We differ from Rad\'o's original and other's further works in that we do not use an extra halting state $H$. For example let us consider the winner Turing machine with 4 states from one of the above mentioned summary paper by \citep{michel-2009}. Here the rule $(2,{\bf 0}) \rightarrow (H, {\bf 1},\rightarrow)$ was applied at last step. So the Rad\'o Busy Beaver fuction $\Sigma$ for $n=4$ is $13$ in contrast with our case, where it is equal only to $12$.   
To summarize we may write that 
\[
\begin{aligned}
    \Sigma(1) &= BB(1) = 1, \\
    \Sigma(2) &= BB(2) = 4, \\
    \Sigma(3) &= BB(3) = 6, \\
\Sigma(4) - 1 &= BB(4) = 12 \text{ and} \\
\Sigma(5) - 1 &= BB(5) \ge 4097.
\end{aligned}
\]
\end{remark}

\begin{example}
In this example in Figure \ref{bb4m}, we show machines which are computed by our brute force $BB_4$ program. We may make an interesting observation that, any two sequential machines are very similar to each other.
We remark that, at the same time, the lengths of these machines (apart from trivial ones) are good estimations of the shortest programs that are computing the 
Busy Beaver 
words from $n=1$ to $n=12$.   

\begin{figure}[htp]
\centering
\subfigure[1,"1",1,1,1]{\label{bb4m1}\includegraphics[scale=0.7]{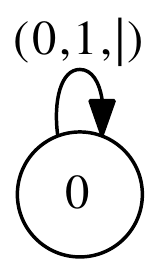}}\qquad
\subfigure[2,"11",2,2,2]{\label{bb4m2}\includegraphics[scale=0.7]{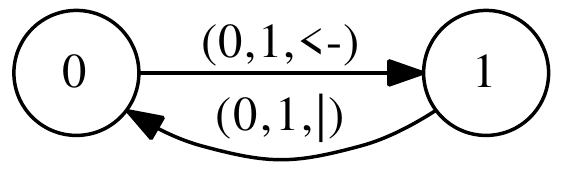}}\qquad
\subfigure[3,"111",2,3,5]{\label{bb4m3}\includegraphics[scale=0.7]{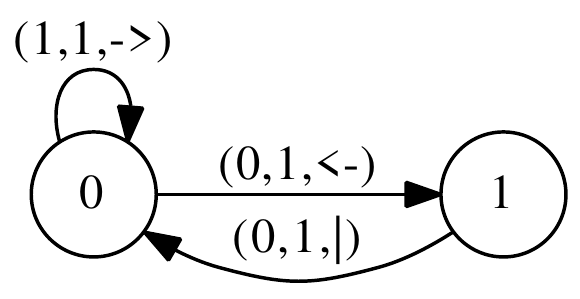}}\\
\subfigure[4,"1111",2,3,5]{\label{bb4m4}\includegraphics[scale=0.7]{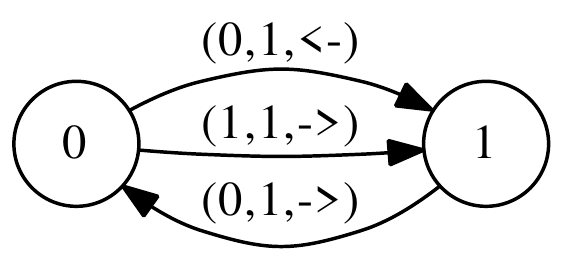}}\qquad
\subfigure[5,"11111",3,4,6]{\label{bb4m5}\includegraphics[scale=0.7]{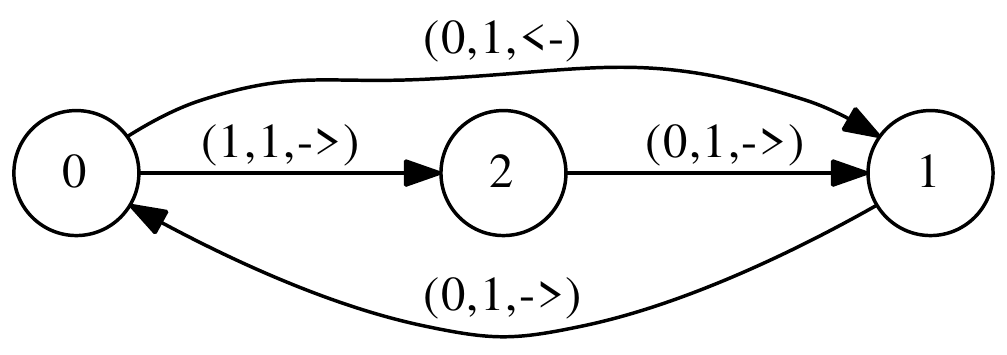}}\\
\subfigure[6,"111111",3,5,16]{\label{bb4m6}\includegraphics[scale=0.5]{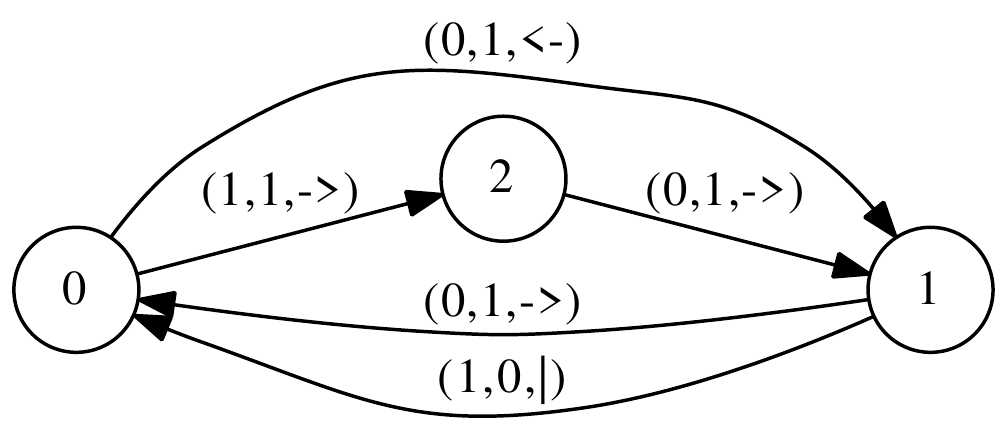}}
\subfigure[7,"1111111",4,6,21]{\label{bb4m7}\includegraphics[scale=0.5]{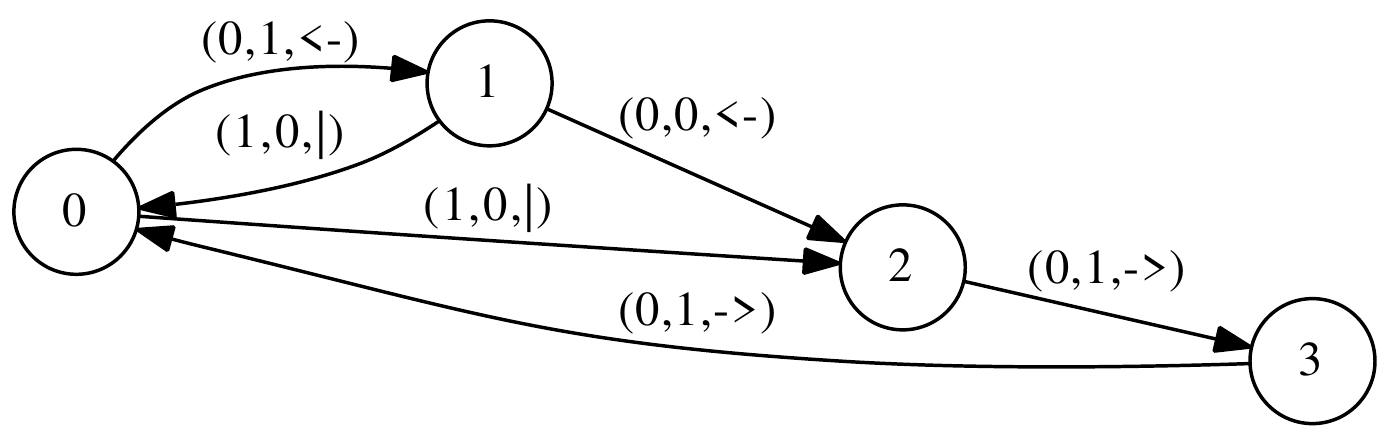}}\\
\subfigure[8, "1010111111",4,6,21]{\label{bb4m8}\includegraphics[scale=0.43]{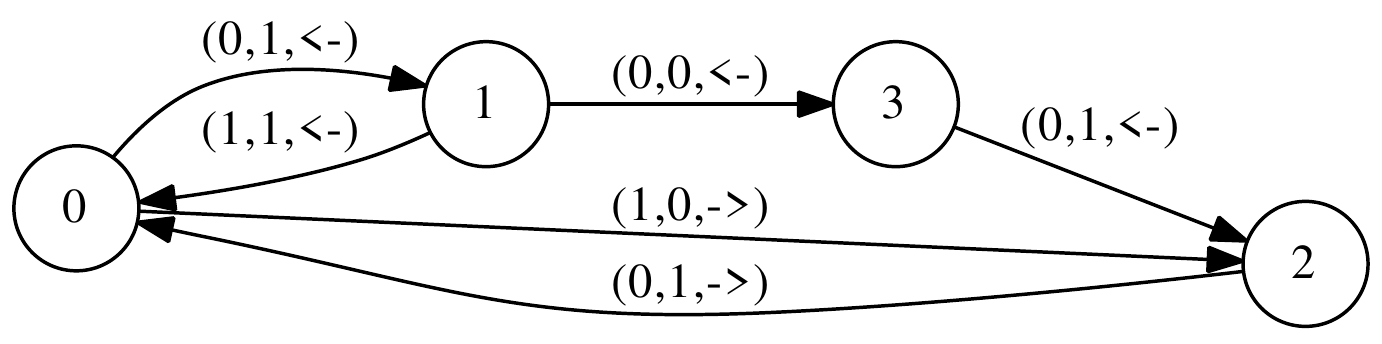}}
\subfigure[9,"111111111",4,6,25]{\label{bb4m9}\includegraphics[scale=0.43]{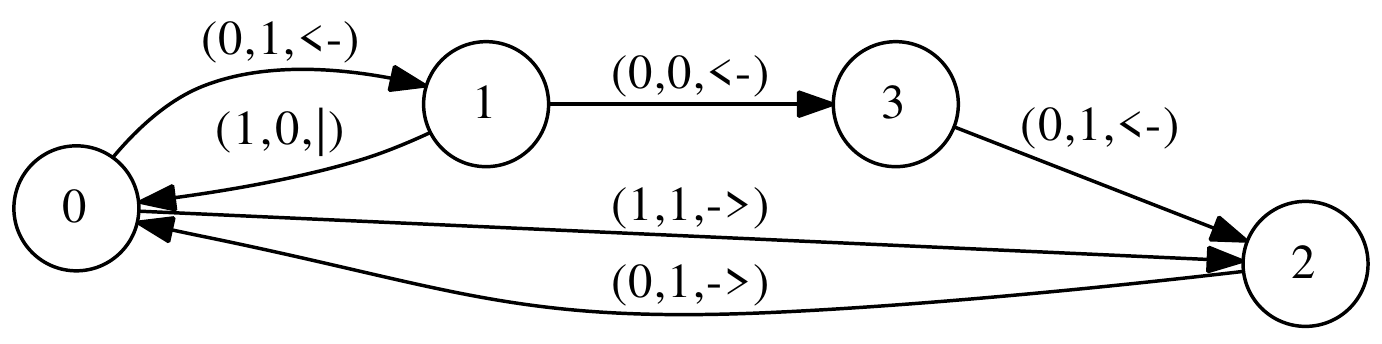}}\\
\subfigure[10,"11111101111",4,7,63]{\label{bb4m10}\includegraphics[scale=0.43]{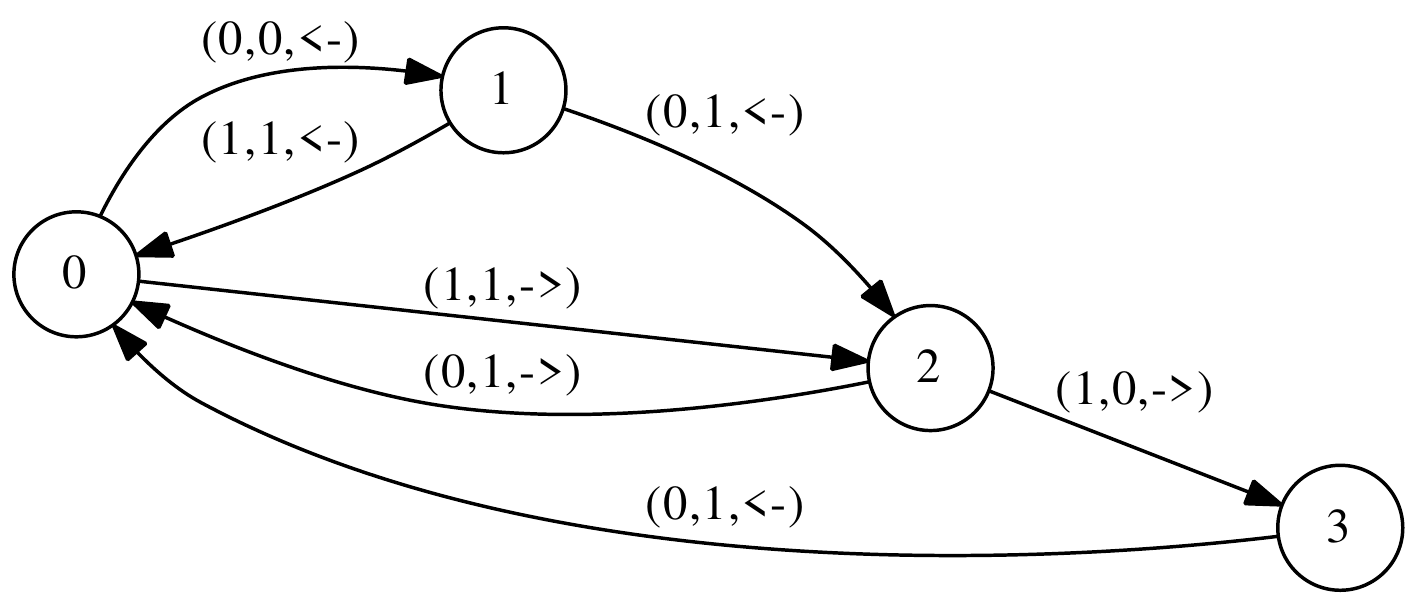}}
\subfigure[11, "$1^{11}$",4,7,45]{\label{bb4m11}\includegraphics[scale=0.43]{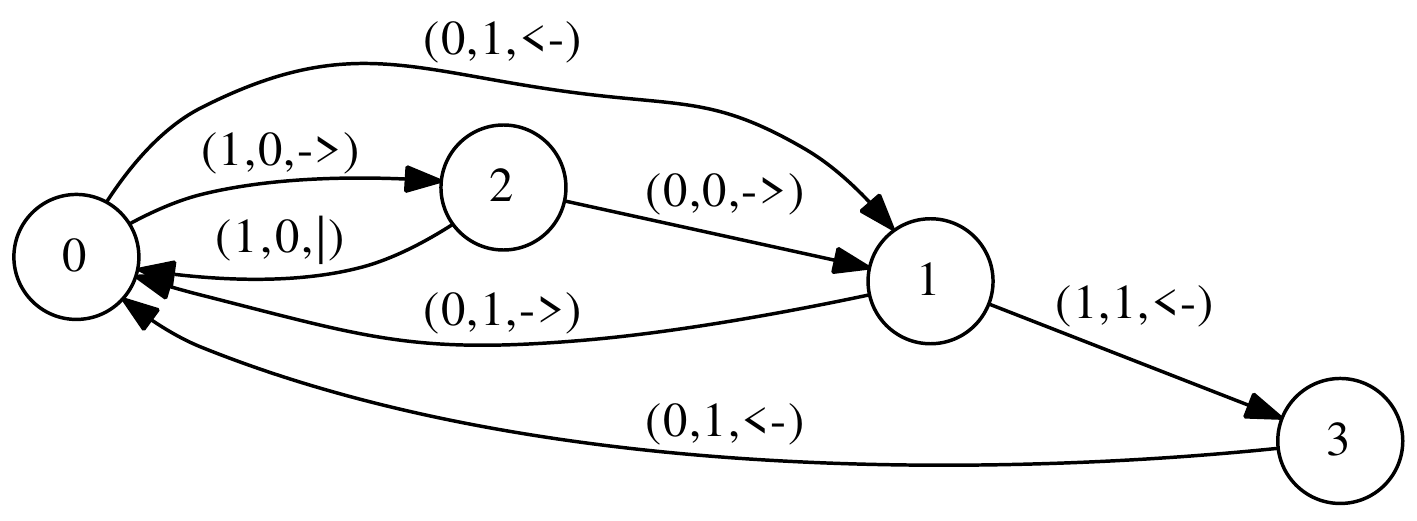}}\\
\subfigure[12, "$1^{12}$",4,7,106]{\label{bb4m12}\includegraphics[scale=0.7]{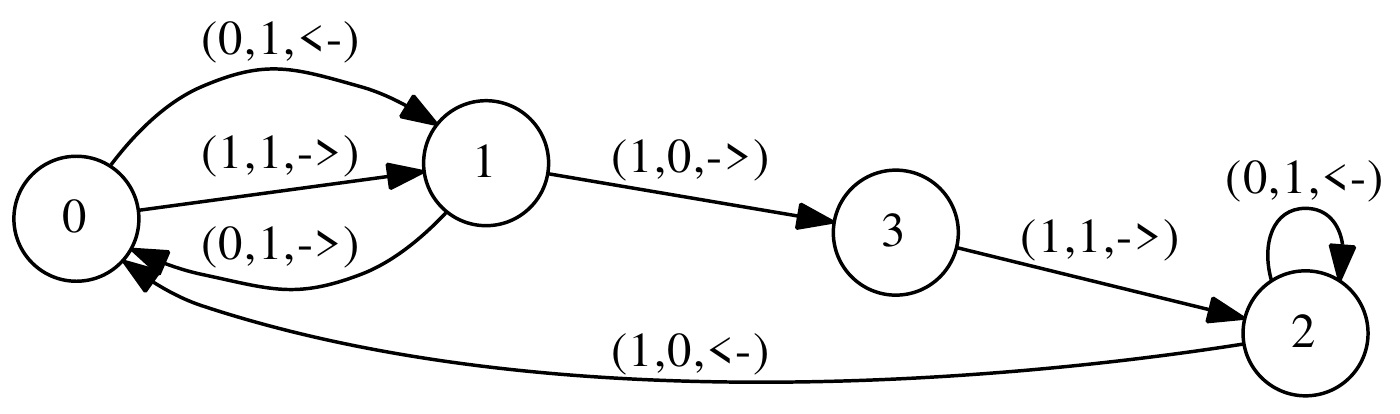}}
\caption{"Placid Platypus machines" are found by our programs, (\# of 1's, "$T(\lambda)$", \# of states, \# of rules, \# of steps)}
\label{bb4m}
\end{figure}

These presented machines are solutions to Placid Platypus problem for $n=1$ to $n=12$.
The Placid Platypus function, introduced by Harland \citep{PlacidPlatypus}, is defined as 
$PP(n) = min\{k \big| \text{nr\_ones}(T(\lambda))=n, T \in BB_k \}$.

\[
\overbrace{
  \overbrace{
    \overbrace{
      \overbrace{\text{ 1 }}^{BB_1}
      \text{ 2 3 4 }}^{BB_2}
    \text{ 5 6 }}^{BB_3}
  \text{ 7 8 9 10 11 12 }}^{BB_4}
\]

$PP(n)= k$, where $k$ such that $BB(k-1) < n \le BB(k)$, (n=1 to 12). The values of Placid Platypus function for some higher $n$ can be found in Harland's works. The questions come easy, but no answers come easy, for example, is the PP function monotonic?
\end{example}

\begin{definition}[Busy Beaver language]\label{BBL}
Let $\mathcal{L}_{BB}$ be the set of Busy Beaver words or in other words, 
the set 
$$\mathcal{L}_{BB} = \{w \in \{0,1\}^* | w \text{ is a Busy Beaver word}\}$$
will be called the Busy Beaver language. 
To be more precise let
$\mathcal{L}_{BB}(n) =  \{w=T(\lambda) | T \in BB_n \text{ and } \text{nr\_ones}(w) = BB(n)\}$ 
then the Busy Beaver language is defined as 
$\mathcal{L}_{BB} = \bigcup_{n}{\mathcal{L}_{BB}(n)}$.

In a similar manner we may define the Busy Beaver Machines as follows
$\mathcal{M}_{BB}(n) = \{T \in BB_n | \text{nr\_ones}(T(\lambda)) = BB(n)\}$
or rather the Placid Platypus Machines as follows
$\mathcal{M}_{PP}(n) = \{T \in BB_k | k = PP(n), \text{nr\_ones}(T(\lambda)) = n\}$.
\end{definition}

\begin{example}
We have carried out our ''brute force'' computations and got the following results.
\[
\begin{aligned}
     \mathcal{L}_{BB} = 
     & \{1\}  \cup \\
     & \{1111\}  \cup \\
     & \{111111, 1011111, 1111101\}  \cup \\
     & \{111111111111\}  \cup \dots
\end{aligned}
\]

\[
\begin{aligned}
     Card(\mathcal{M}_{BB}(1)) &= 1, \\ 
     Card(\mathcal{M}_{BB}(2)) &= 2, \\ 
     Card(\mathcal{M}_{BB}(3)) &= 44, \\ 
     Card(\mathcal{M}_{BB}(4)) &= 24, \dots
\end{aligned}
\]

$K_TBB(n) = min\{t_Z(\lambda) | Z \in \mathcal{M}_{BB}(n)\}$

\[
\begin{aligned}
    &K_TBB(1) = 1, \\
    &K_TBB(2) = 5, \\
    &K_TBB(3) = 10, \\
    &K_TBB(4) = 95,  \dots 
\end{aligned}
\]
\end{example}

In the following, some theorems on Busy Beaver Machines are stated without proofs, but we give some examples. These simple theorems are useful in processing results of our searching programs which are, for example, shown in appendixes. 

\begin{theorem}[Operating symmetry]
Let $T$ and $T'$ be two Turing machines, $T'$ obtained from $T$ by reversing to opposite direction of movement. Then 
$T'(\lambda)$ is the mirror word of $T(\lambda)$ and $t_{T'}(\lambda) = t_{T}(\lambda)$.
\end{theorem}

\begin{example}
For example, the two machines shown in Figure  \ref{mirror1} and \ref{mirror2} or \ref{mbb4m1} and \ref{mbb4m3} are symmetrical in operating.
\end{example}

\begin{theorem}[State symmetry]
Let $T$ and $T'$ be two Turing machines, $T'$ obtained from $T$ by commuting states differ from $0$. Then 
$T(\lambda) = T'(\lambda)$ and $t_{T}(\lambda) = t_{T'}(\lambda)$.
\end{theorem}

\begin{example}
For example, machines in Figure \ref{mbb4m1} and \ref{mbb4m5} are symmetrical in states.
\end{example}

\begin{definition}
Call a subset $\mathcal{B}_{BB}(n)$ of $\mathcal{M}_{BB}(n)$ base machines if it has the following property: any machine $T$ in the $\mathcal{M}_{BB}(n)$ can be represented as operating or state symmetry of a base machine.
\end{definition}

\begin{example}
A base machines set for $\mathcal{M}_{BB}(4)$ is shown in the following figure.
\begin{figure}[htp]
\centering
\subfigure["$1^{12}$",106]{\includegraphics[scale=0.43]{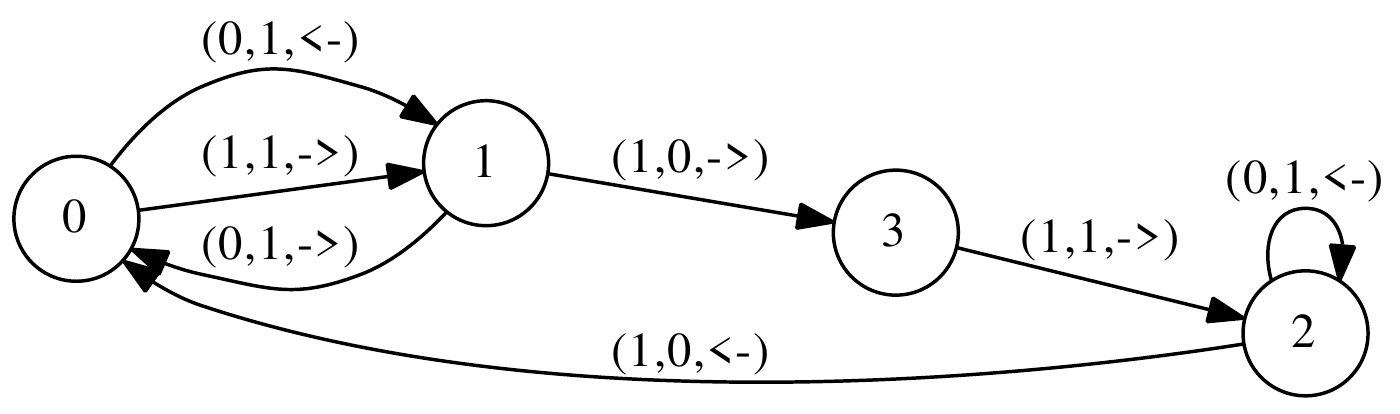}} %
\subfigure["$1^{12}$",95]{\includegraphics[scale=0.43]{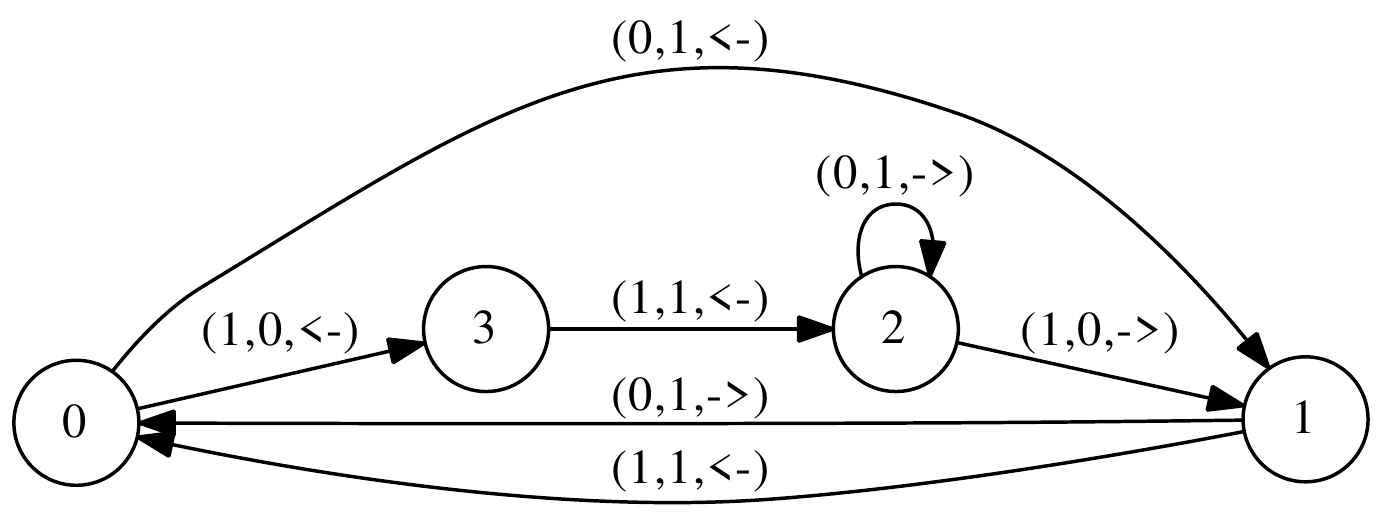}} %
\caption{A base machines set for $\mathcal{M}_{BB}(4)$.}
\end{figure}
All machines of $\mathcal{M}_{BB}(4)$ are shown in Appendix \ref{mbb4}. All of them can be represented with these two base machines.

The elements of a base machines set for $\mathcal{M}_{BB}(3)$ are shown in the following figures: \ref{mbb3m1},\ref{mbb3m2},\ref{mbb3m3},\ref{mbb3m4},\ref{mbb3m5},\ref{mbb3m6},\ref{mbb3m7},\ref{mbb3m8},\ref{mbb3m9},\ref{mbb3m10} and \ref{mbb3m11}.

Cardinalities of known base machines can be summarized as follows.
\[
\begin{aligned}
     Card(\mathcal{B}_{BB}(1)) &= 1, \\ 
     Card(\mathcal{B}_{BB}(2)) &= 1, \\ 
     Card(\mathcal{B}_{BB}(3)) &= 11, \\ 
     Card(\mathcal{B}_{BB}(4)) &= 2
\end{aligned}
\]
\end{example}

\subsection{Running Time of the Shortest Programs}

It is well known, for example from \citep{Chaitin}, that the length of Busy Beaver words grows faster than any computable function, accordingly $K_T(w)$ grows at least fast than $|w|$, because $|w| \le K_T(w)$, so thus we should try to search such estimations, which are based on $K_T(w)/|w|$ rather than $K_T(w)$.   

Let $w$ be a Busy Beaver word. The $|w|=\text{nr\_ones}(w)$, $K(w)=PP(\text{nr\_ones}(w))$ and $K_T(w)=\text{Rad\'o's } S(|w|)$ approximations are used in the following table.

\begin{longtable}[h]{|c|c|c|c|c|}
\label{runningtimes}
$\text{nr\_ones}(w)$&	
$\frac{K_T(w)}{|w|}$&
$\frac{|w|}{K(w)}$&
$\frac{K_T(w)K(w)}{|w|^2}$&
Source\\

\hline
\endhead

1&	1,0&	1,0&	1,0& own programs\\
\hline
2&	1,0&	1,0&	1,0& own programs\\
\hline
3&	1,6666&	1,5&	1,1111& own programs\\
\hline
4&	1,2500&	2,0&	0,625& own programs\\
\hline
5&	1,2&	1,6666&	0,72& own programs\\
\hline
6&	2,6666&	2,0&	1,3333& own programs\\
\hline
7&	3,0&	1,75&	1,7142& own programs\\
\hline
8&	2,625&	2,0&	1,3125& own programs\\
\hline
9&	2,7777&	2,25&	1,2345& own programs\\
\hline
10&	6,3&	2,5&	2,52& own programs\\
\hline
11&	4,0909&	2,75&	1,4876& own programs\\
\hline
\textbf{12}&	8,8333&	3,0&	2,9444& own programs\\
\hline
\hline
13&	8,2307&	2,6&	3,1656& own programs\\
\hline
14&	22,3571&	2,8&	7,9846& own programs\\
\hline
15&	7,3333&	3,0&	2,4444& own programs\\
\hline
16&	12,4375&	3,2&	3,8867& own programs\\
\hline
17&	8,7058&	3,4&	2,5605& own programs\\
\hline
18&	6,2222&	3,6&	1,7283& own programs\\
\hline
19&	10,1052&	3,8&	2,6592& own programs\\
\hline
21&	24,4761&	4,2&	5,8276& own programs\\
\hline
26&	10,1153&	5,2&	1,9452& own programs\\
\hline
32&	18,1875&	6,4&	2,8417& own programs\\
\hline
117&	115,282&	23,4&	4,9265& Brady\\
\hline
160&	130,7937&	32&	4,0873& own programs\\
\hline
501&	268,3972&	100,2&	2,6786& Schult\\
\hline
1915&	1114,095&	383&	2,9088& Uhing\\
\hline
4097&	2879,8711&	819,4&	3,5146& recombinated\\
\hline
4097&	2880,6155&	819,4&	3,5155& recombinated\\
\hline
4097&	2881,3634&	819,4&	3,5164& recombinated\\
\hline
4097&	4317,5618&	819,4&	5,2691& recombinated\\
\hline
4097&	4317,5652&	819,4&	5,2691& recombinated\\
\hline
4097&	5756,0004&	819,4&	7,0246& recombinated\\
\hline
\textbf{4097}&	11514,979&	819,4&	14,0529& Marxen, Buntrock\\
\hline
4097&	17266,4898&	819,4&	21,0721& recombinated\\
\hline
\hline
136612&	96057,2482&	22768,6666&	4,2188& Marxen, Buntrock\\
\hline
95524079&	90975317,14&	15920679,83&	5,7142& Marxen, Buntrock\\
\hline
2,5e+21&	2,12e21&	5e+20&	5,088& Marxen, Buntrock\\
\hline
6.4e+462&	9,53125e+462&	1,0666e+462&	8,9361& Marxen, Buntrock\\
\hline
1,2e+865&	2,5e+865&	2e+864&	12,5& Marxen, Buntrock\\
\hline
2,5e+881&	3,56e+881&	4,16e+880&	8,544& T. and S. Ligocki\\
\hline
\textbf{4,6e+1439}&	5,4e+1439&	7,6e+1438&	7,0434& T. and S. Ligocki\\
\hline

\caption{The increasing of running time of the shortest programs.}

\end{longtable}

The not our own $\text{Rad\'o's } S(|w|)$ and $\Sigma(|w|)$ values contained in this table can be found in \citep{michel-2009}. The recombinated machines can be found in \citep{batfai-recomb-2009}.

The values of 4th column in the table (namely that $K_T(w)=O(|w|^2)$ encourage us to believe that there may exists an estimation of running time of the shortest programs.

\section{Conclusion and further work}
In this paper we have posed a question, namely, how long does it take to run the shortest programs? Cannot the shortest programs run on indefinitely in time? Our intuition and first calculations shown in table \ref{runningtimes} both suggest an answer in the affirmative.

We focused on Busy Beaver machines rather than general case of the shortest Turing machines. 
Busy Beaver machines and words have been charted fully for $n=1$ to $4$ with our programs. We have already begun programs to chart the Busy Beaver case of $n=5$ and Placid Platypus machines. Our searching program in principle may find such $BB_5$ machines that can do more than 135 million steps before halting. But our main purpose is to produce much more $\frac{K_T(w)K(w)}{|w|^2}$ values.

\appendix

\clearpage
\section{Placid Platypus Machines}
In this appendix we show some $\mathcal{M}_{PP}(n)$ machines where $PP(n) = 5$. 
Some Placid Platypus Machines for $PP(n) = 1 \text{ to } 4$ were already shown in Figure \ref{bb4m}. 

The following machines, in order of increasing $n$, are also founded by our C programs. These programs and related data can be downloaded from \url{http://www.inf.unideb.hu/~nbatfai/bb}.

\subsection{$\mathcal{M}_{PP}(n), PP(n)=5$}

\begin{figure}[htp]
\centering
\subfigure[13, "$11111111111101$",5,8,107]{\label{pp13}\includegraphics[scale=0.38]{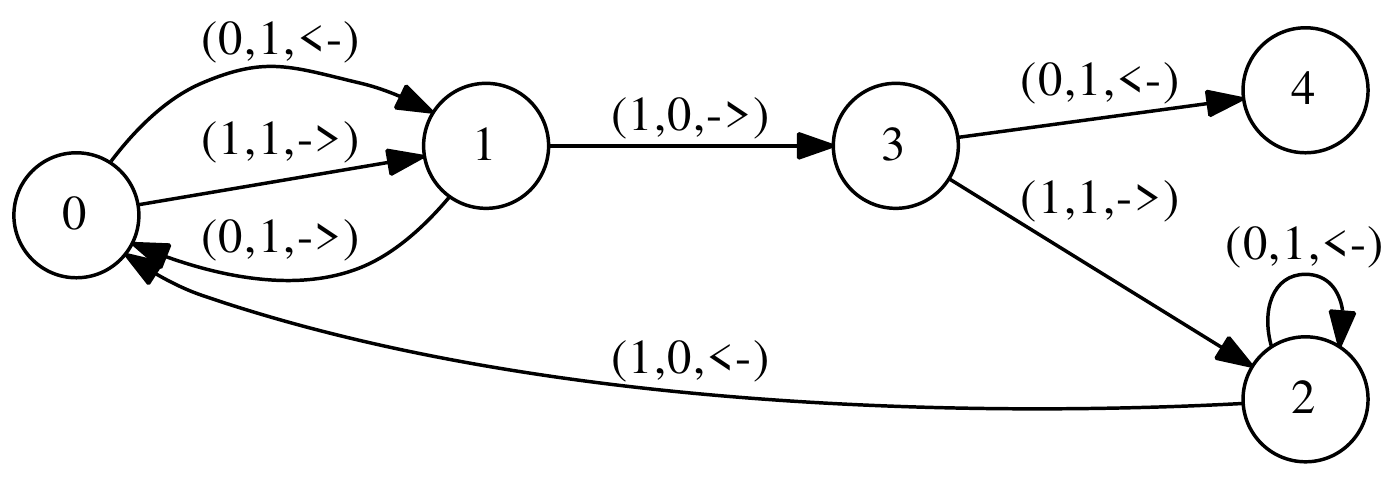}} % BB5/BB5-8-f4w-4f.txt (hpserver)
\subfigure[14, "$1^{12}00101$",5,9,313]{\label{pp14}\includegraphics[scale=0.38]{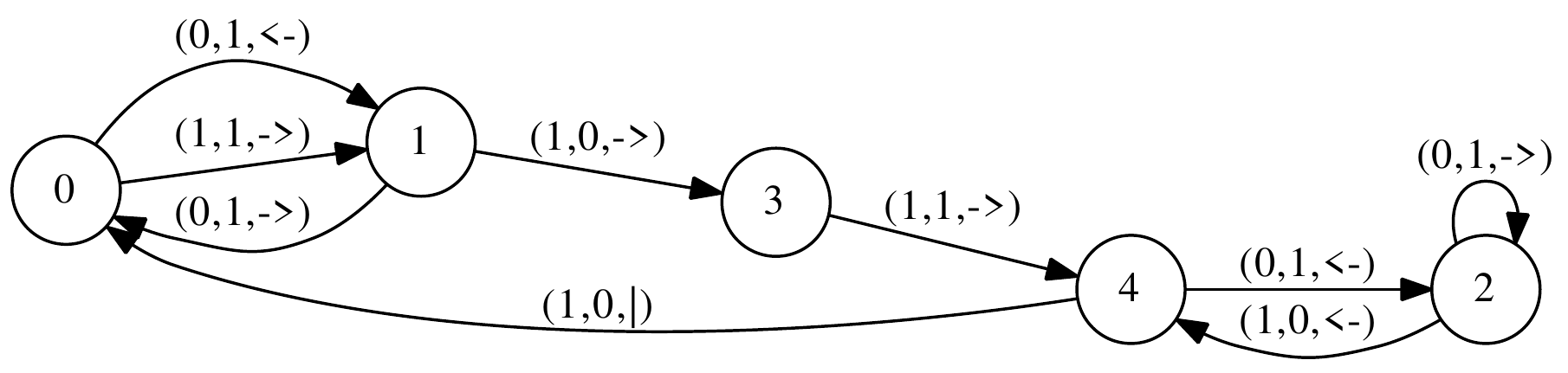}} \\% BB5/BB5-9-f4w.txt (mz)
\subfigure[15, "$1^{15}$",5,9,110]{\label{pp15}\includegraphics[scale=0.38]{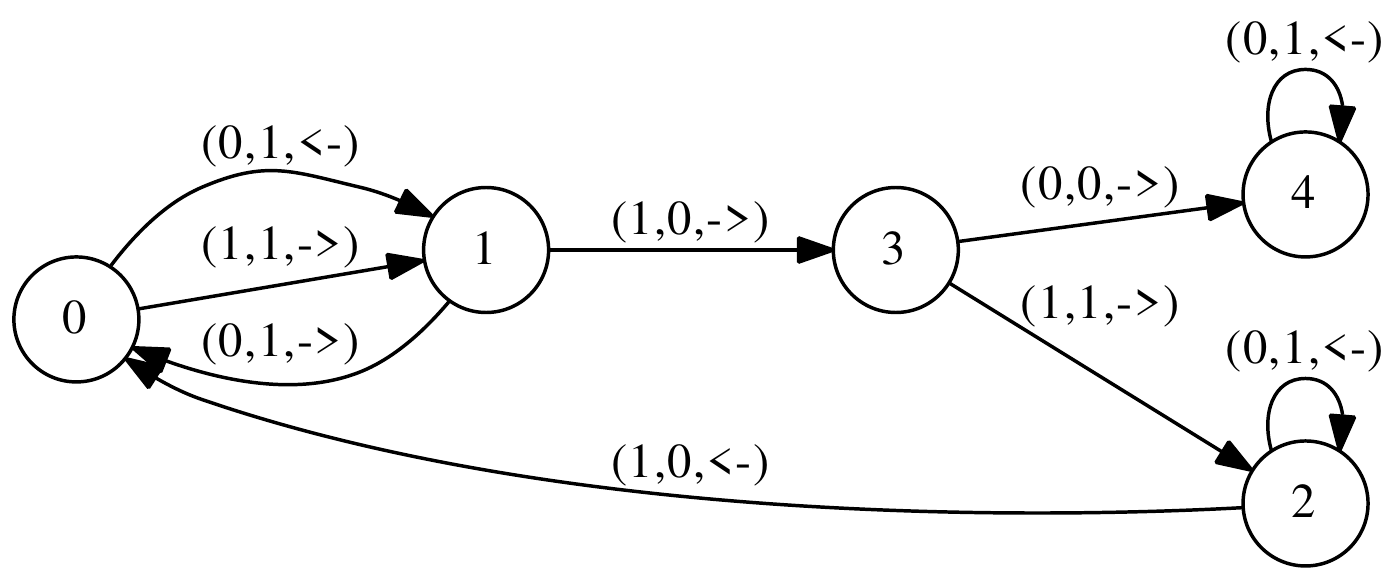}} % BB5/BB5-9-f4w-5f.txt (mz)
\subfigure[16, "$111101^{8}01111$",5,9,199]{\label{pp16}\includegraphics[scale=0.38]{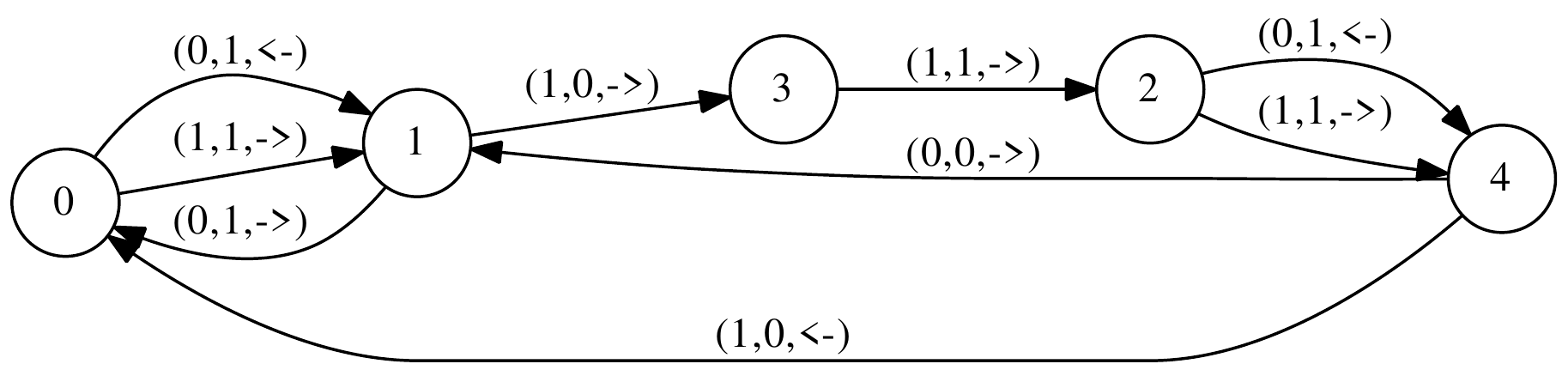}} \\% BB5/BB5-9-f4w.txt (mz)
\subfigure[17, "$101^{16}$",5,9,148]{\label{pp17}\includegraphics[scale=0.33]{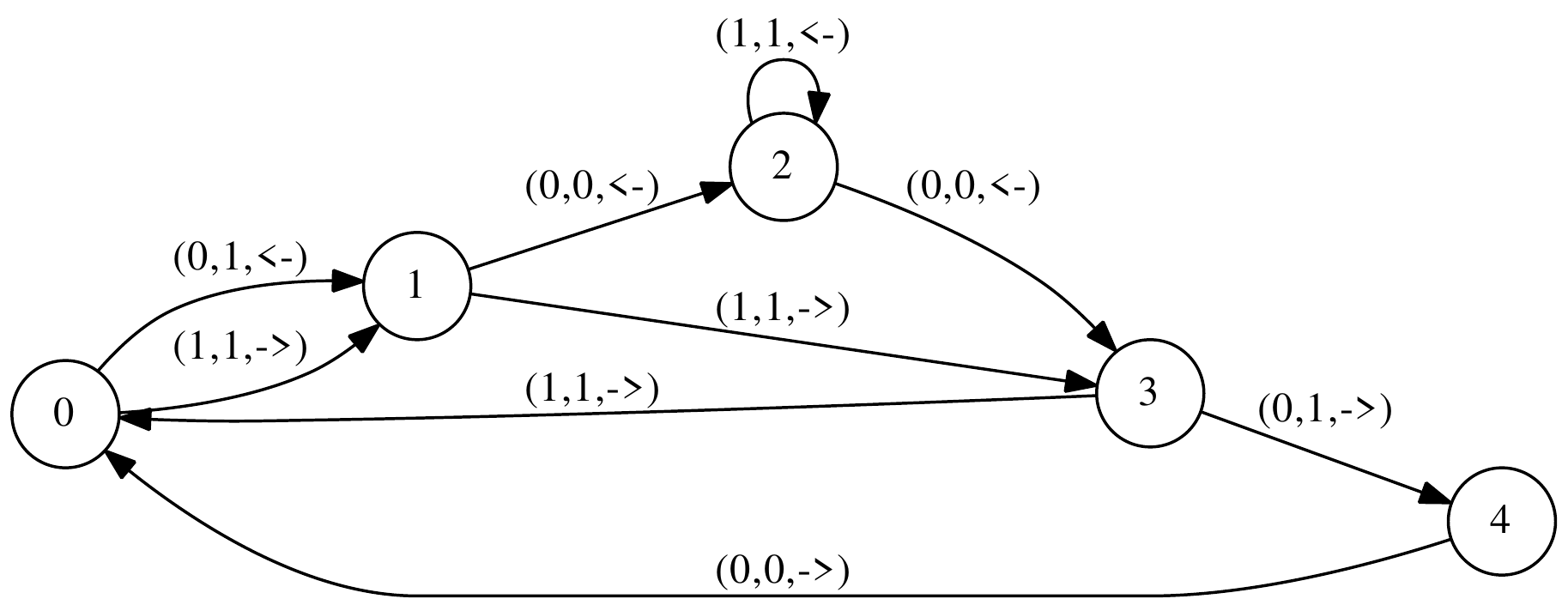}} % BB5/BB5-9-f4w-5f.txt (mz)
\subfigure[18, "$1^{10}01^{8}$",5,9,112]{\label{pp18}\includegraphics[scale=0.33]{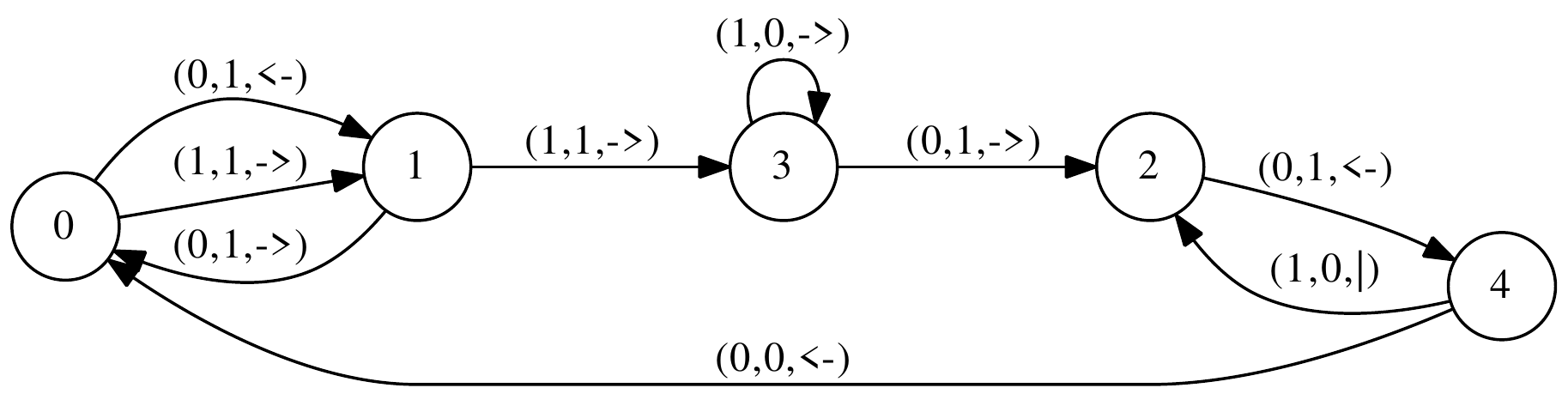}} % BB2/5/bb5ms9-exp5.txt (hpserver)
\caption{Some Placid Platypus machines are found by our programs, (\# of 1's, "$T(\lambda)$", \# of states, \# of rules, \# of steps)}
\end{figure}

\begin{figure}[htp]
\centering
\subfigure[19, "$111101^{7}01110111011$",5,9,192]{\label{pp19}\includegraphics[scale=0.38]{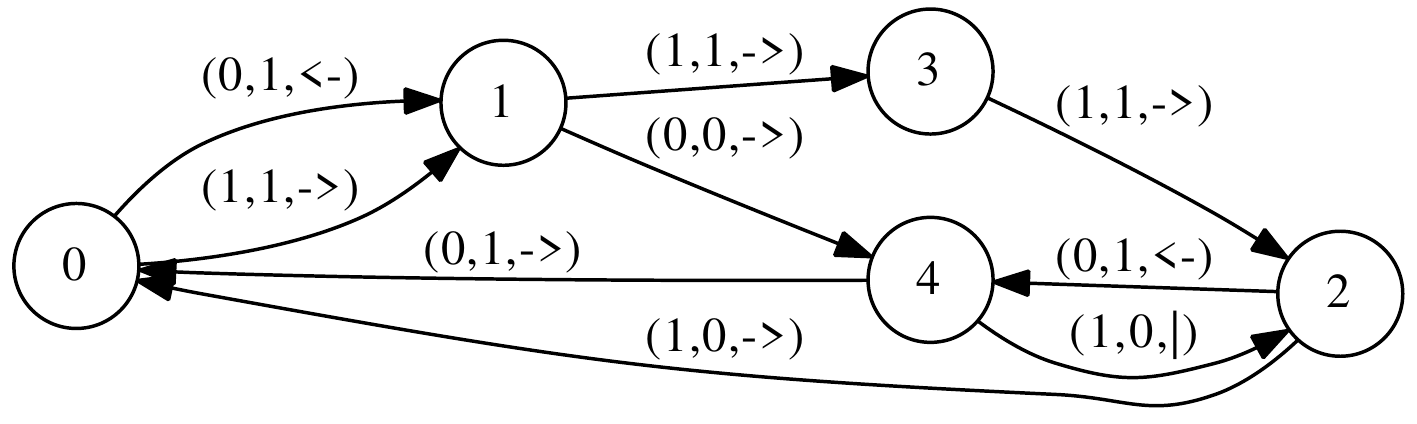}} % BB5/BB5-9-f4w.txt (mz)
\subfigure[21, "$(101001)^{2}10101^{11}011$",5,9,514]{\label{pp21}\includegraphics[scale=0.38]{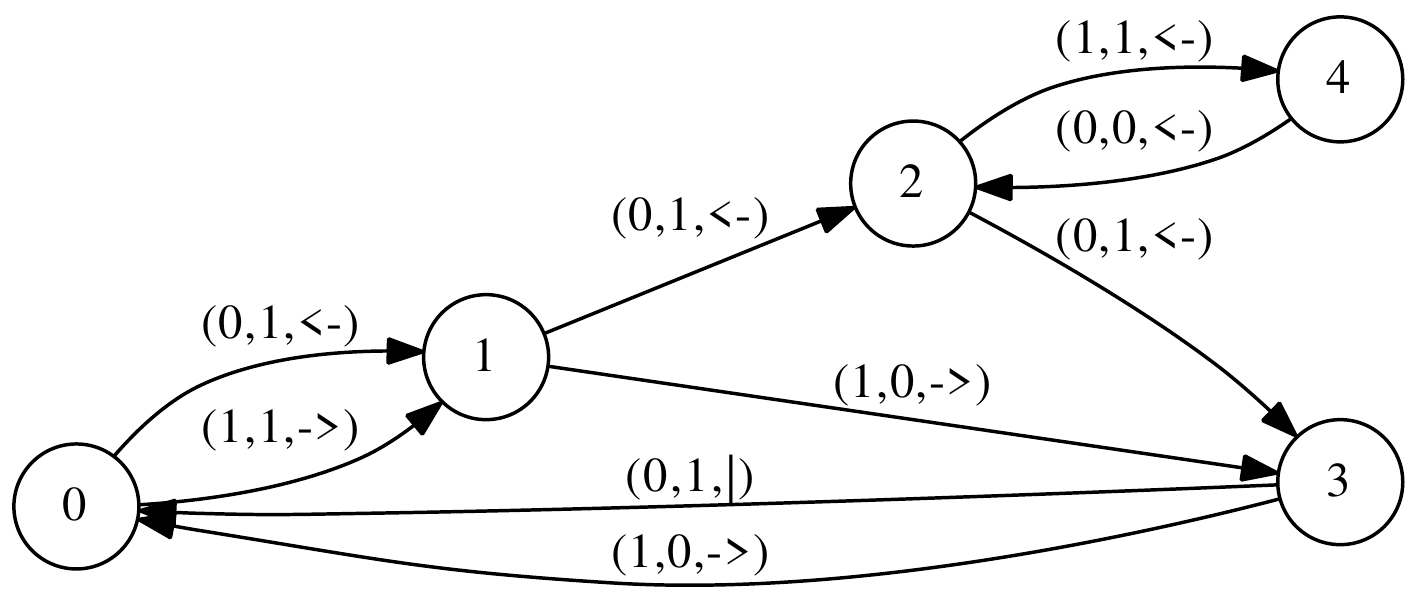}} \\% ./BB5-9-f4w-5f
\subfigure[26, "$1^{26}$",5,9,263]{\label{pp26}\includegraphics[scale=0.38]{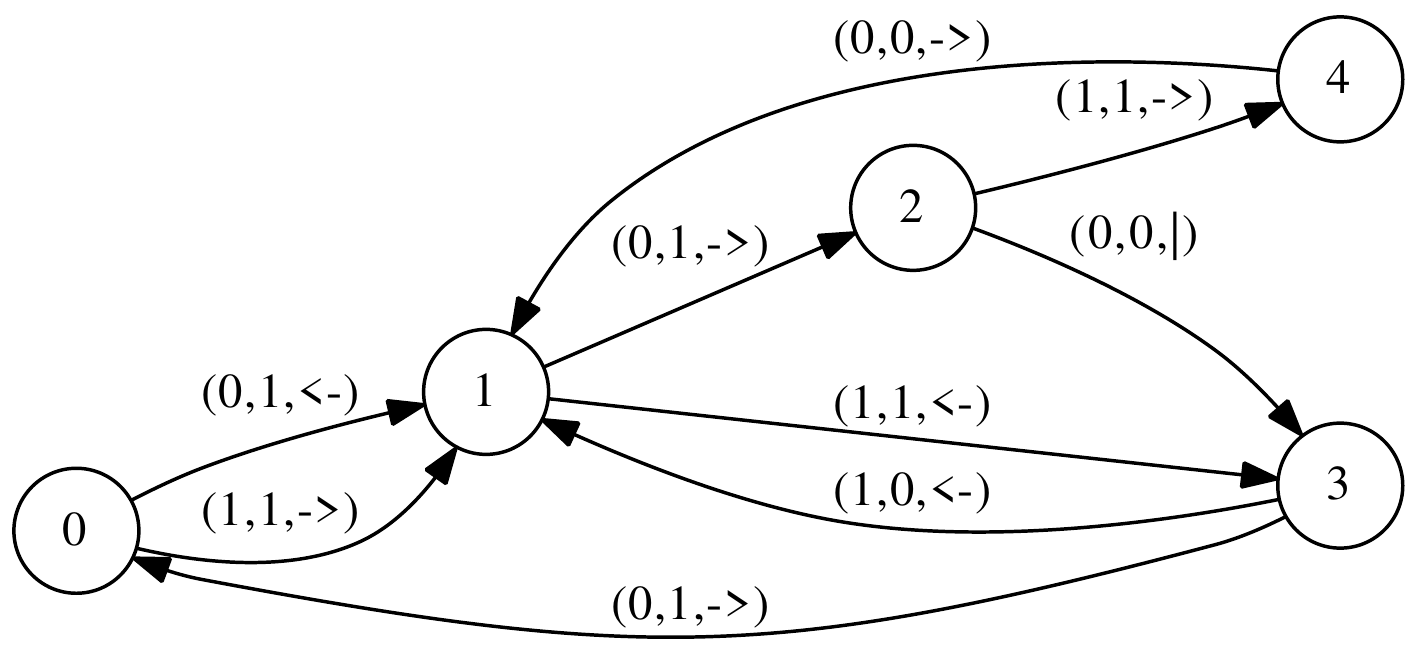}} % ./BB5-9-f4w-5f
\subfigure[32, "$(1110)^{10}11$",5,9,582]{\label{pp32}\includegraphics[scale=0.38]{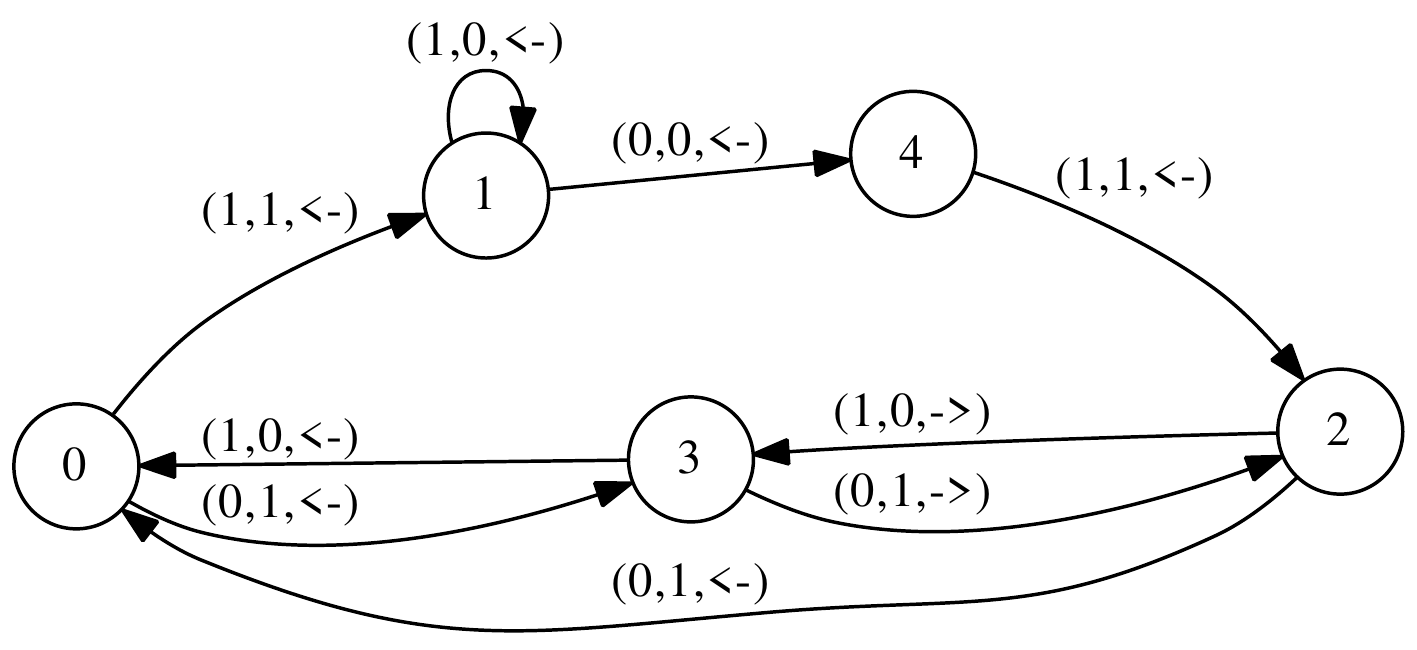}} % ./BB5-9-EXP
\caption{Some Placid Platypus machines are found by our programs, (\# of 1's, "$T(\lambda)$", \# of states, \# of rules, \# of steps)}
\end{figure}

\begin{figure}[htp]
\centering
\subfigure[160, "$1^{160}$",5,9,20927]{\label{pp160}\includegraphics[scale=0.43]{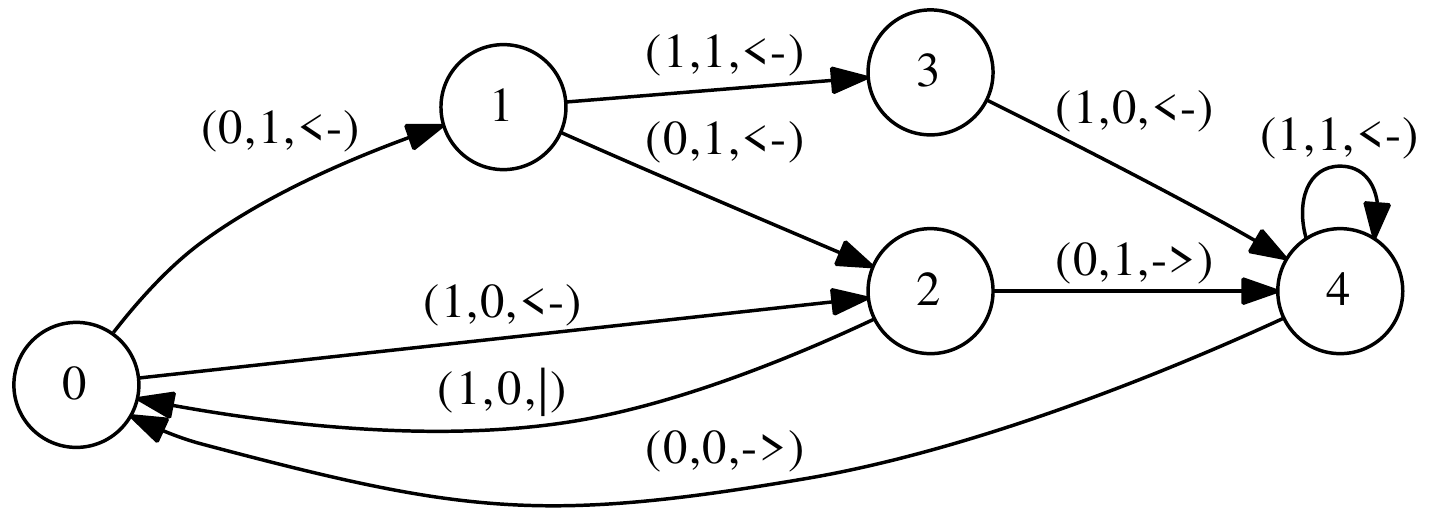}} % ./BB5-9-f4w
\caption{Some Placid Platypus machines are found by our programs, (\# of 1's, "$T(\lambda)$", \# of states, \# of rules, \# of steps)}
\end{figure}

\clearpage
\section{Busy Beaver Machines}
In this appendix we show all $\mathcal{M}_{BB}(n)$ machines for $n=1 \text{ to } 4$. 

The following machines, in order of found, are also computed by our C programs. These programs and related data can be downloaded from \url{http://www.inf.unideb.hu/~nbatfai/bb}.

\subsection{$\mathcal{M}_{BB}(1)$}

\begin{figure}[h!]
\centering
\subfigure["1",1]{\includegraphics[scale=0.7]{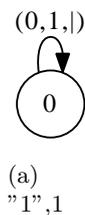}} %
\caption{The only $\mathcal{M}_{BB}(1)$ Busy Beaver Machine, ("$T(\lambda)$", \# of steps)}
\end{figure}

\subsection{$\mathcal{M}_{BB}(2)$}

The two $\mathcal{M}_{BB}(2)$ Busy Beaver Machines are symmetric in operation.

\begin{figure}[h!]
\centering
\subfigure["$1^{4}$",5]{\includegraphics[scale=0.7]{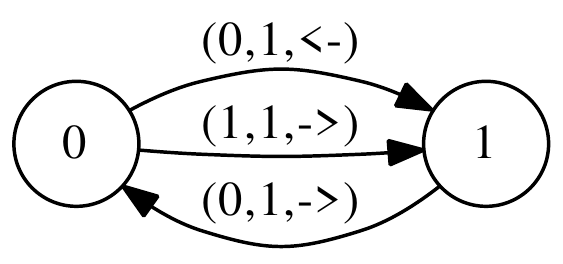}} %
\subfigure["$1^{4}$",5]{\includegraphics[scale=0.7]{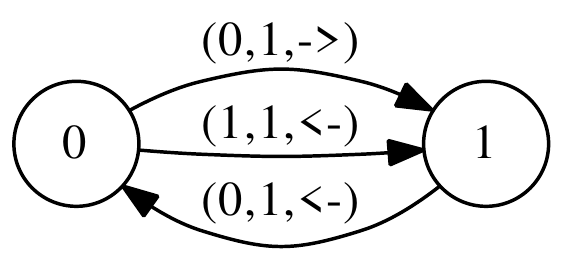}} %
\caption{The two symmetric $\mathcal{M}_{BB}(2)$ Busy Beaver Machines, ("$T(\lambda)$", \# of steps)}
\end{figure}

\clearpage
\subsection{$\mathcal{M}_{BB}(3)$}

\begin{figure}[h!]
\centering
\subfigure["$1^{6}$",16]{\label{mbb3m1}\includegraphics[scale=0.6]{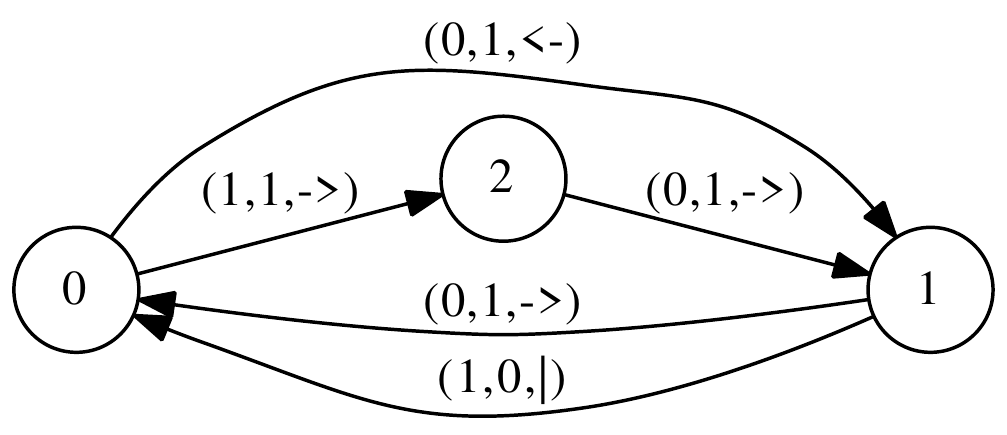}} %
\subfigure["$1^{6}$",12]{\label{mbb3m2}\includegraphics[scale=0.6]{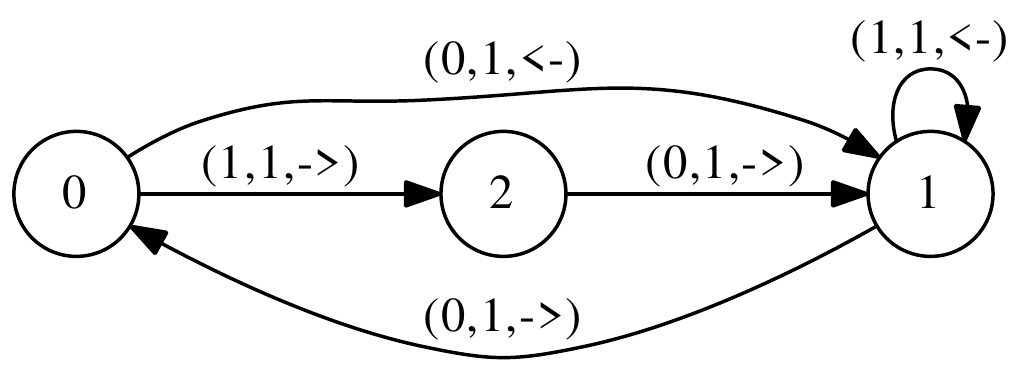}} \\ %
\subfigure["$1^{6}$",14]{\label{mbb3m3}\includegraphics[scale=0.6]{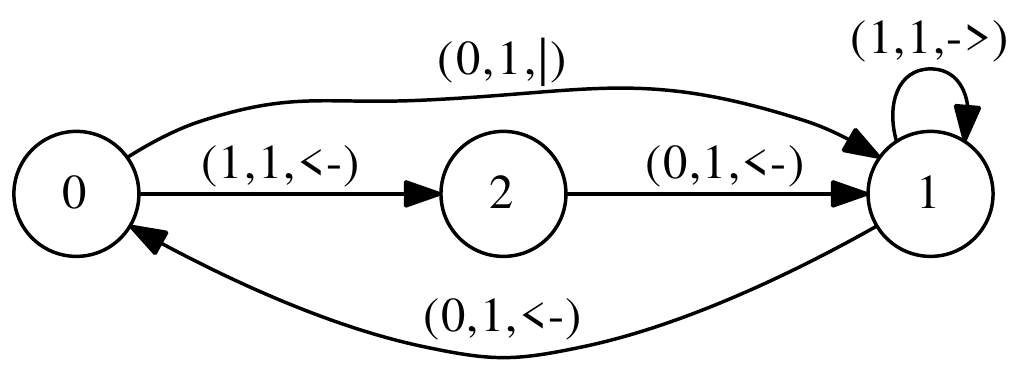}} %
\subfigure["$1^{6}$",14]{\includegraphics[scale=0.6]{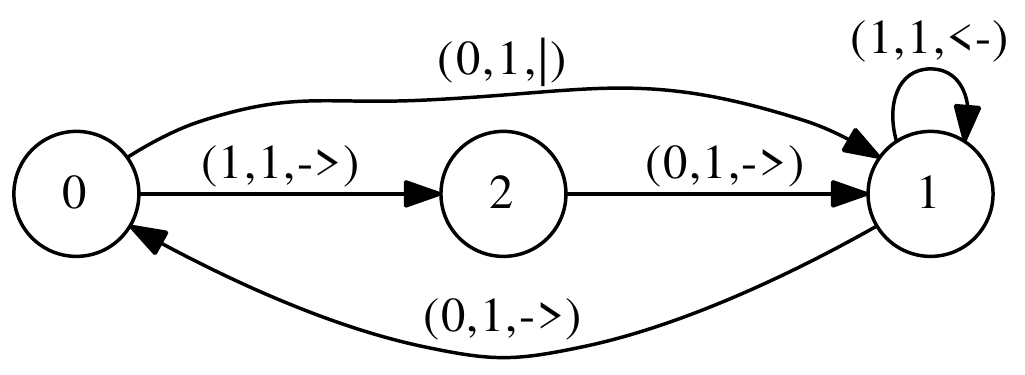}} \\ %
\subfigure["$1^{6}$",16]{\includegraphics[scale=0.6]{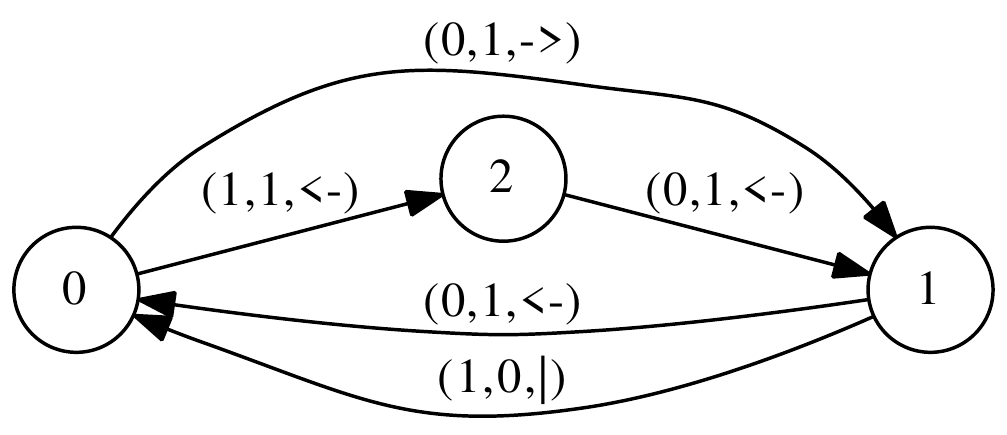}} %
\subfigure["$1^{6}$",12]{\includegraphics[scale=0.6]{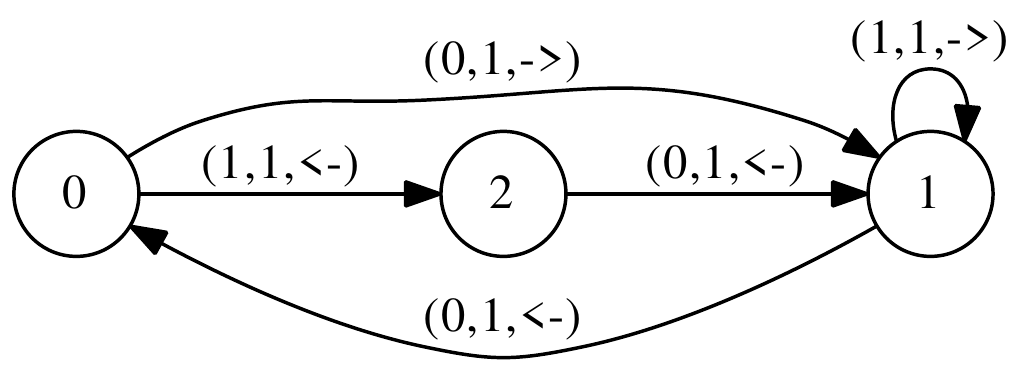}} \\ %
\subfigure["$1^{6}$",11]{\label{mbb3m4}\includegraphics[scale=0.6]{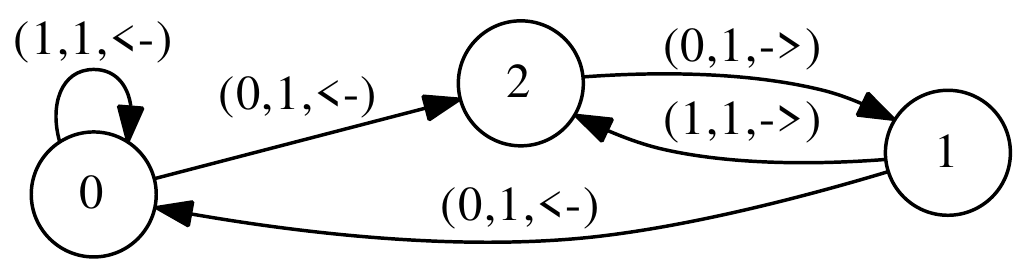}} %
\subfigure["$1^{6}$",12]{\label{mbb3m5}\includegraphics[scale=0.6]{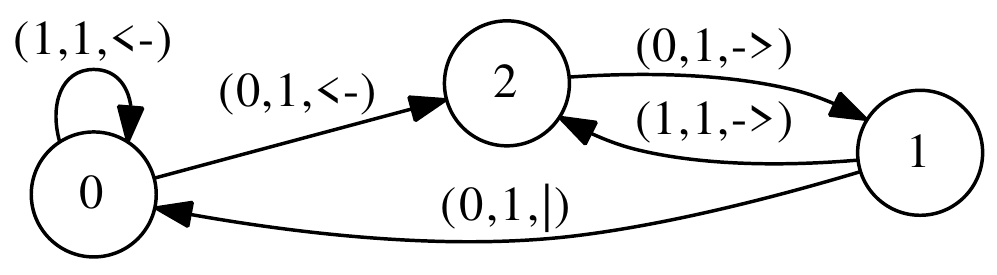}} \\ %
\subfigure["$1^{6}$",14]{\label{mbb3m6}\includegraphics[scale=0.6]{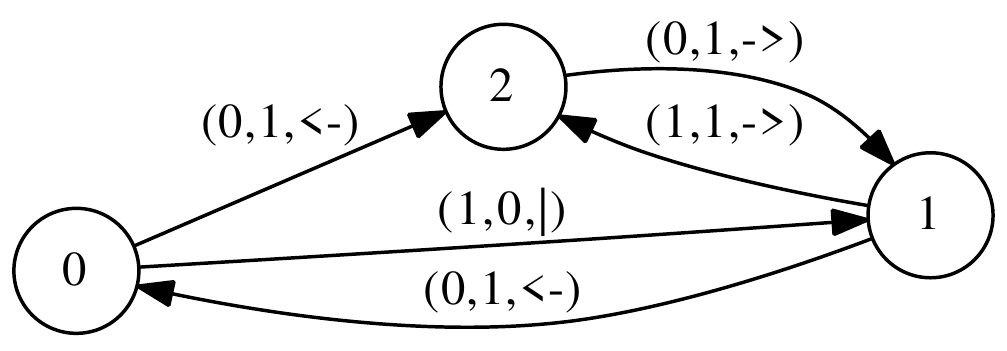}} %
\subfigure["$1011111$",12]{\label{mbb3m7}\label{mirror1}\includegraphics[scale=0.6]{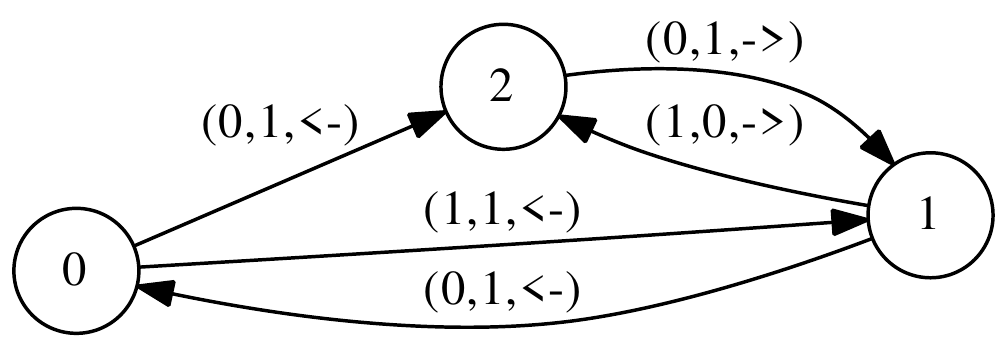}} 
\caption{All $\mathcal{M}_{BB}(3)$ Busy Beaver Machines (0-9), ("$T(\lambda)$", \# of steps)}
\end{figure}

\begin{figure}[h!]
\centering
\subfigure["$1^{6}$",10]{\label{mbb3m8}\includegraphics[scale=0.6]{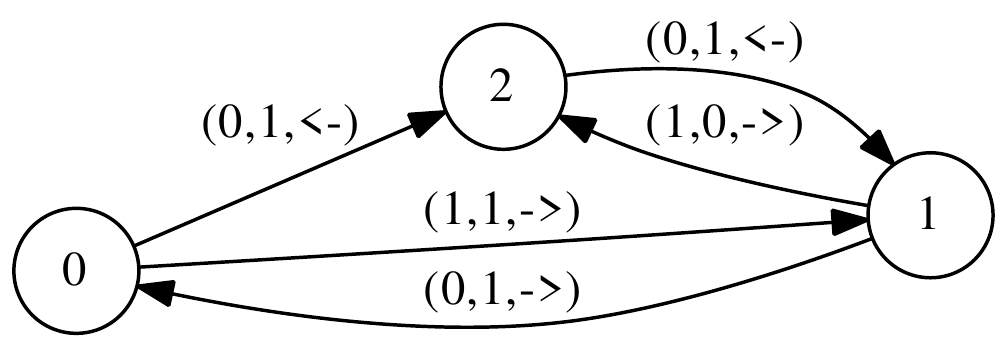}} %
\subfigure["$1^{6}$",12]{\includegraphics[scale=0.6]{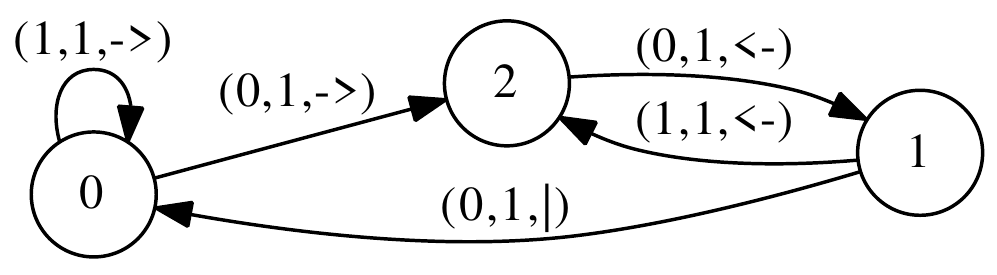}} \\ %
\subfigure["$1^{6}$",11]{\includegraphics[scale=0.6]{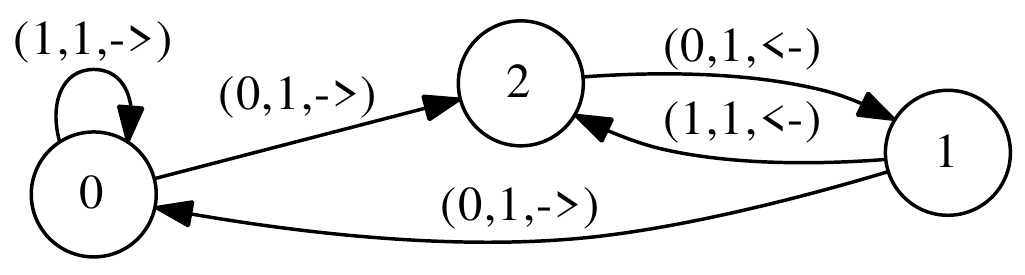}} %
\subfigure["$1^{6}$",14]{\includegraphics[scale=0.6]{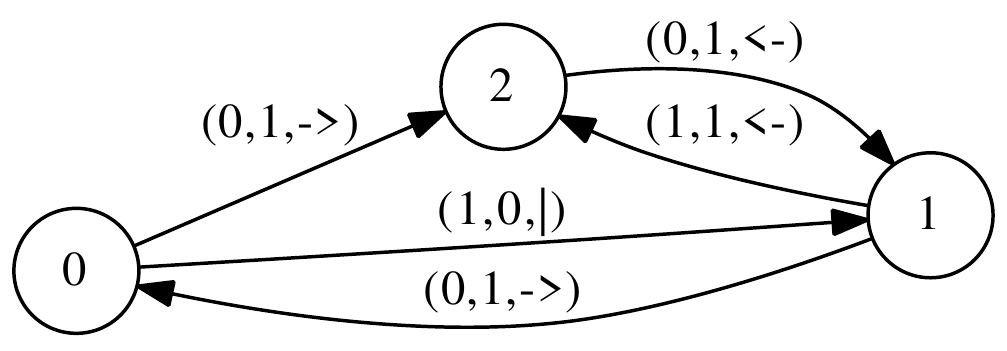}} \\ %
\subfigure["$1^{6}$",10]{\includegraphics[scale=0.6]{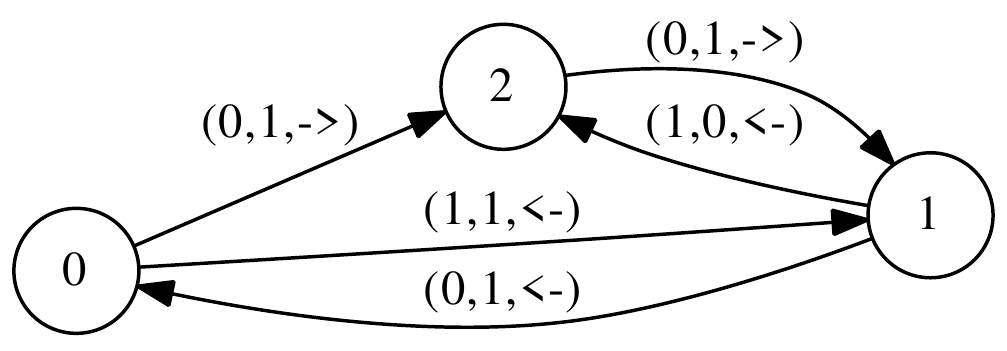}} %
\subfigure["$1111101$",12]{\label{mirror2}\includegraphics[scale=0.6]{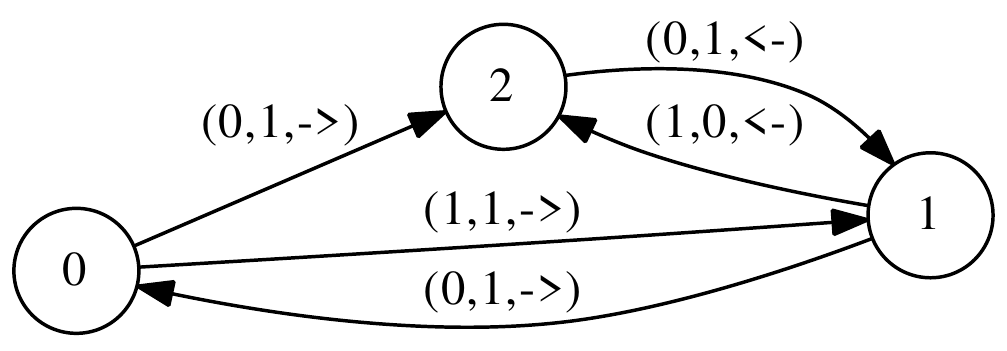}} \\ %
\subfigure["$1^{6}$",11]{\includegraphics[scale=0.6]{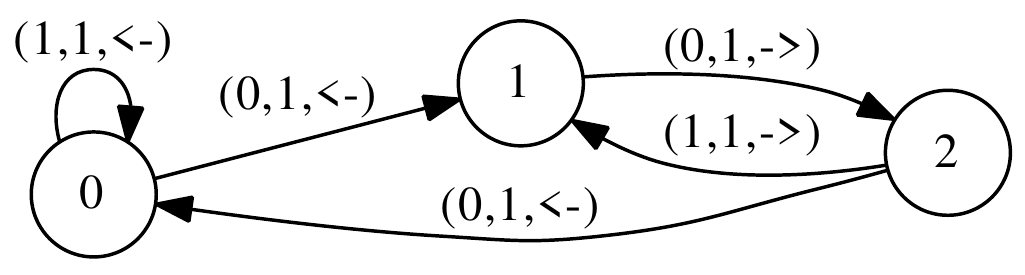}} %
\subfigure["$1^{6}$",12]{\includegraphics[scale=0.6]{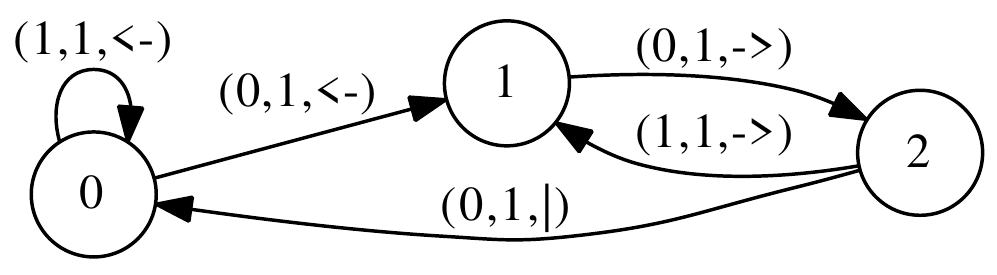}} \\ %
\subfigure["$1^{6}$",14]{\includegraphics[scale=0.6]{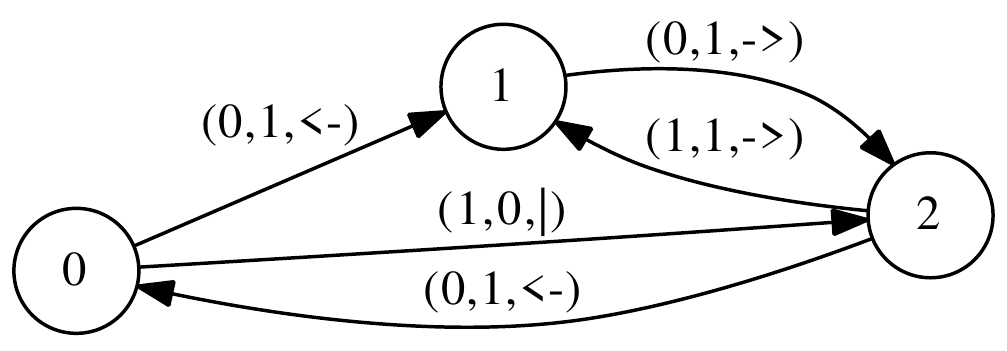}} %
\subfigure["$1011111$",12]{\includegraphics[scale=0.6]{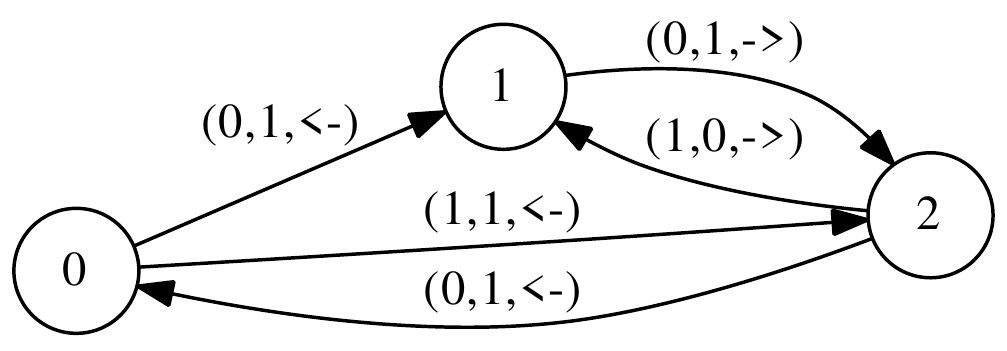}} 
\caption{All $\mathcal{M}_{BB}(3)$ Busy Beaver Machines (10-19), ("$T(\lambda)$", \# of steps)}
\end{figure}

\begin{figure}[h!]
\centering
\subfigure["$1^{6}$",10]{\includegraphics[scale=0.6]{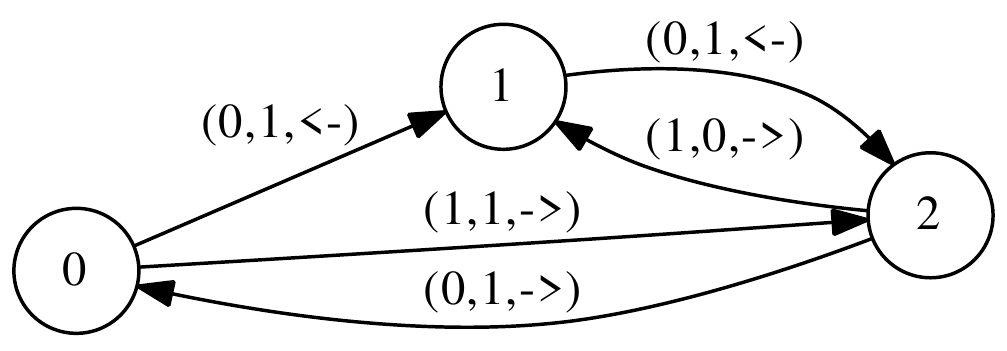}} %
\subfigure["$1^{6}$",12]{\includegraphics[scale=0.6]{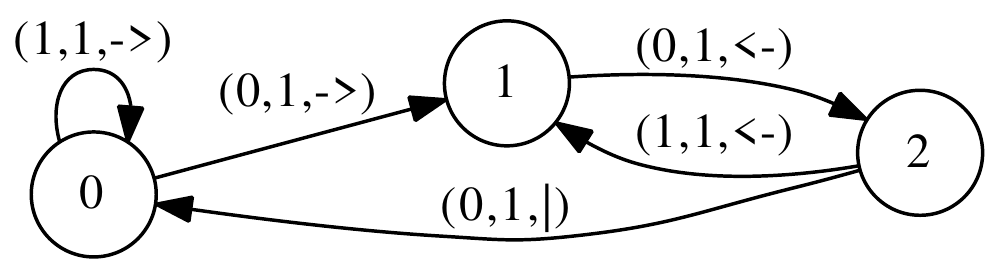}} \\ %
\subfigure["$1^{6}$",11]{\includegraphics[scale=0.6]{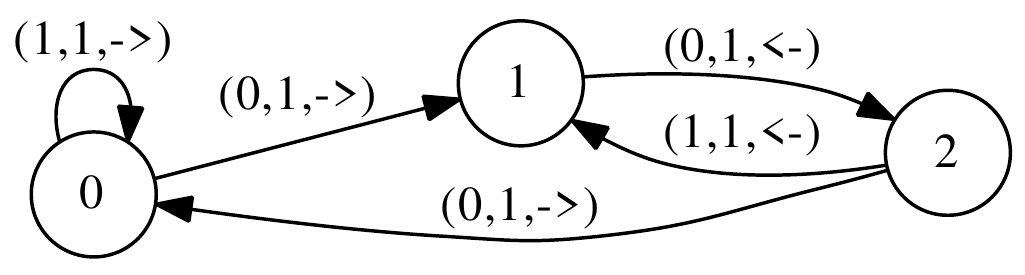}} %
\subfigure["$1^{6}$",14]{\includegraphics[scale=0.6]{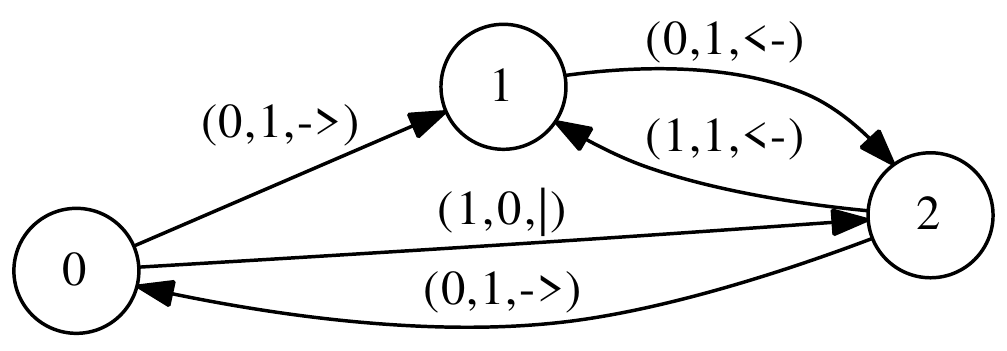}} \\ %
\subfigure["$1^{6}$",10]{\includegraphics[scale=0.6]{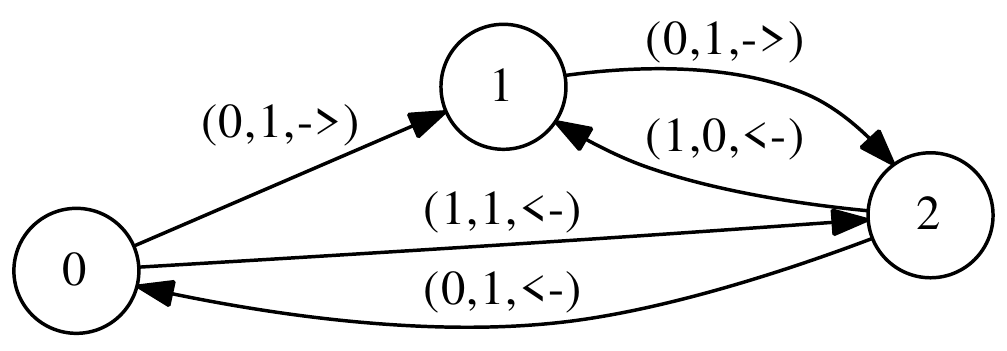}} %
\subfigure["$1111101$",12]{\includegraphics[scale=0.6]{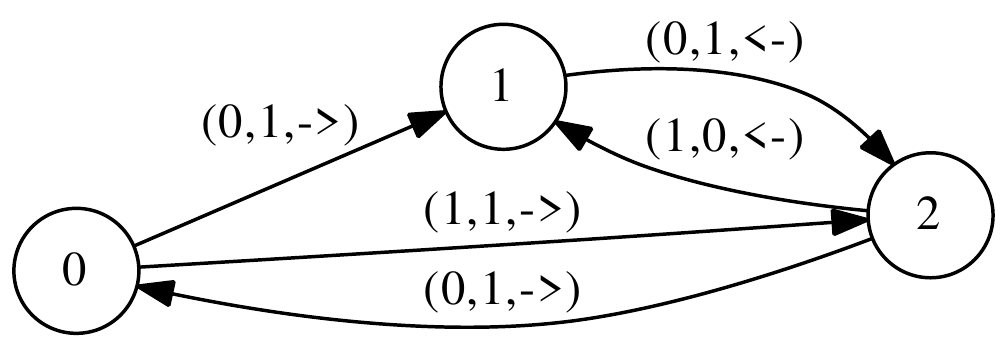}} \\ %
\subfigure["$1^{6}$",16]{\includegraphics[scale=0.6]{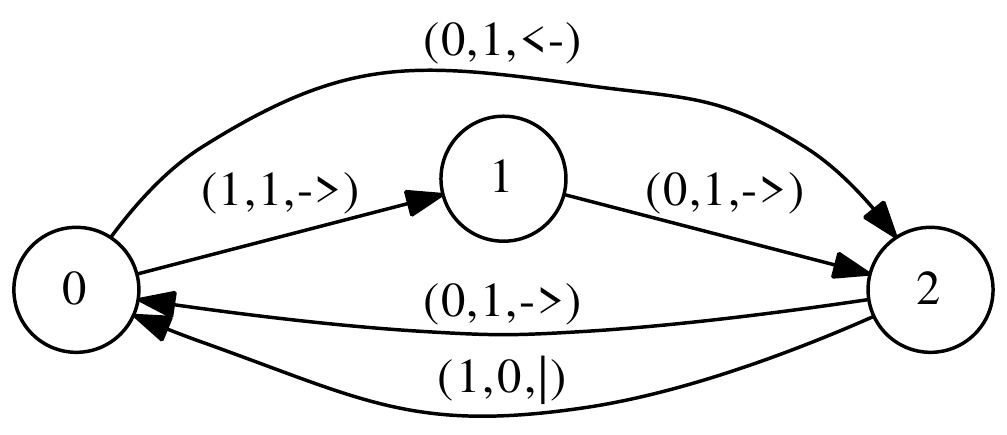}} %
\subfigure["$1^{6}$",12]{\includegraphics[scale=0.6]{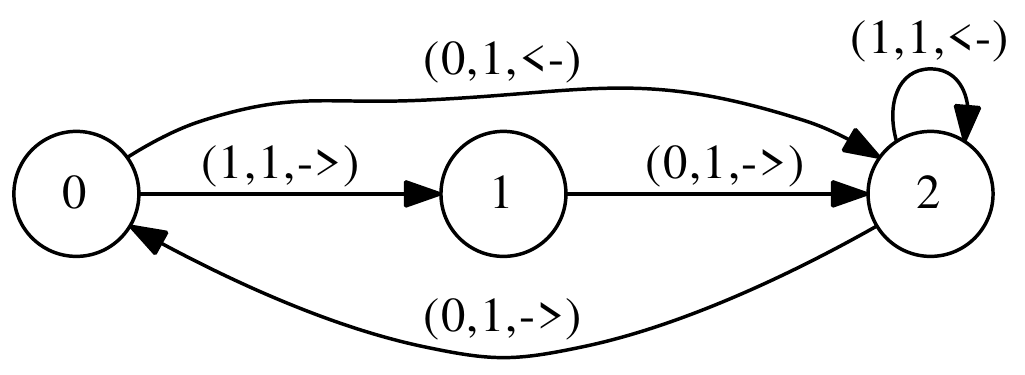}} \\ %
\subfigure["$1^{6}$",14]{\includegraphics[scale=0.6]{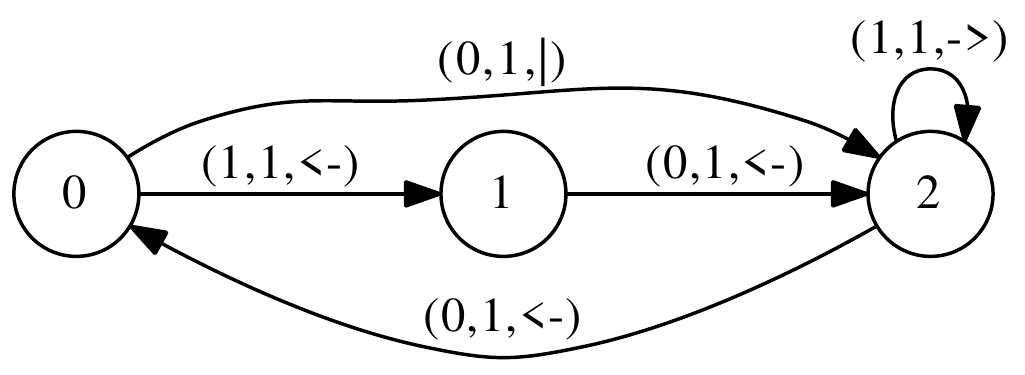}} %
\subfigure["$1^{6}$",14]{\includegraphics[scale=0.6]{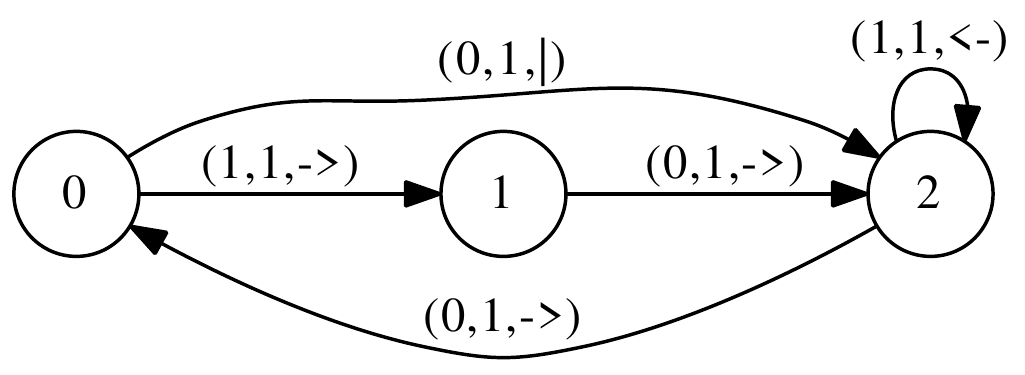}} 
\caption{All $\mathcal{M}_{BB}(3)$ Busy Beaver Machines (20-29), ("$T(\lambda)$", \# of steps)}
\end{figure}

\begin{figure}[h!]
\centering
\subfigure["$1^{6}$",16]{\includegraphics[scale=0.6]{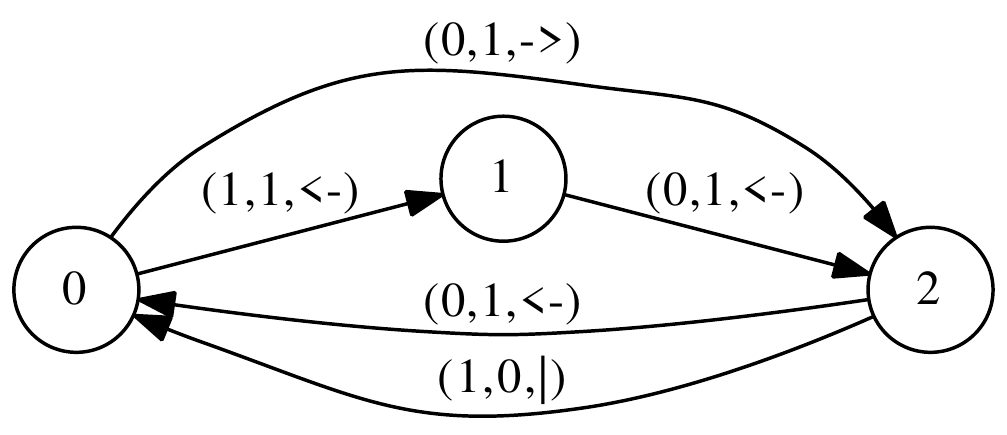}} %
\subfigure["$1^{6}$",12]{\includegraphics[scale=0.6]{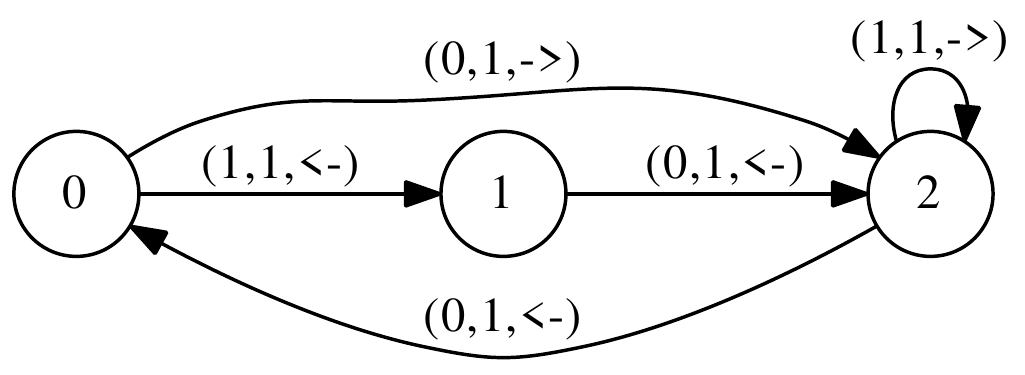}} \\ %
\subfigure["$1^{6}$",16]{\label{mbb3m9}\includegraphics[scale=0.6]{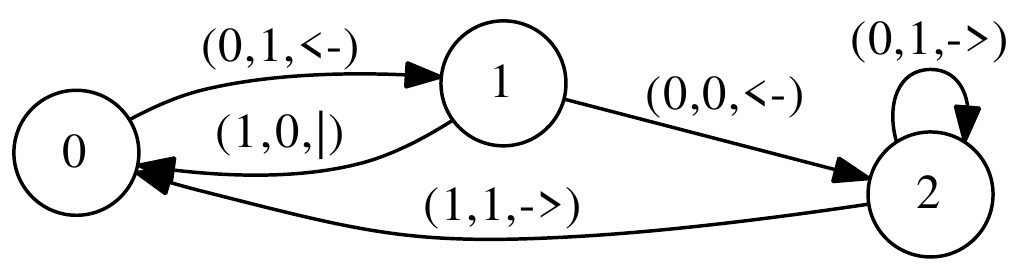}} %
\subfigure["$1^{6}$",13]{\label{mbb3m10}\includegraphics[scale=0.6]{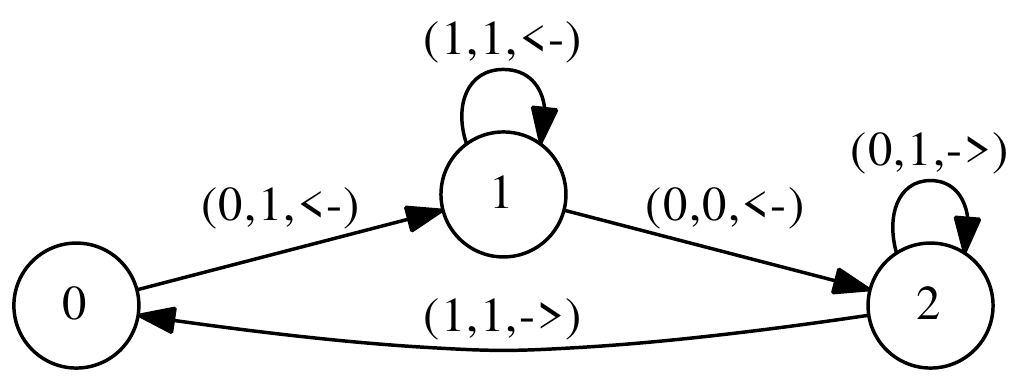}} \\ %
\subfigure["$1^{6}$",15]{\label{mbb3m11}\includegraphics[scale=0.6]{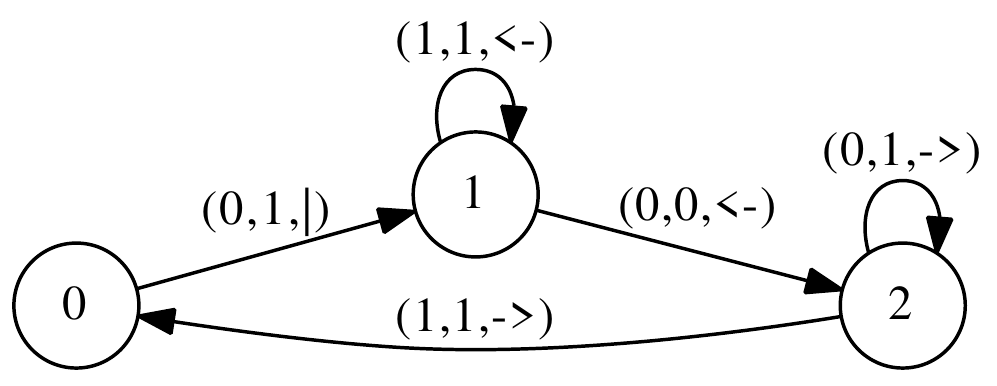}} %
\subfigure["$1^{6}$",15]{\includegraphics[scale=0.6]{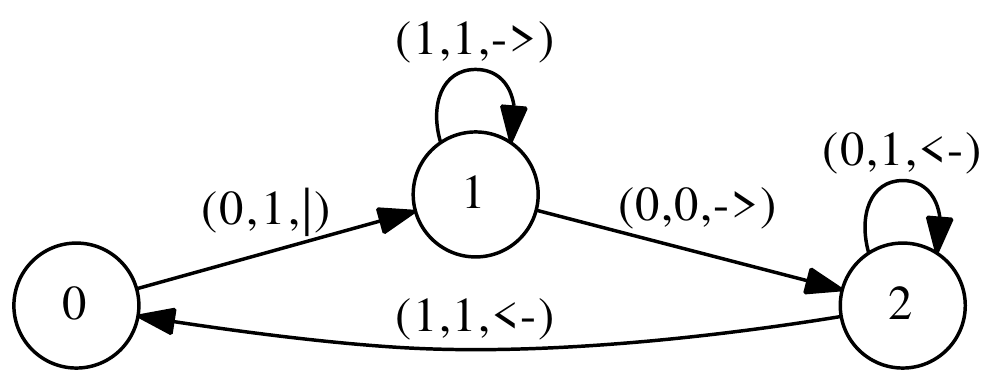}} \\ %
\subfigure["$1^{6}$",16]{\includegraphics[scale=0.6]{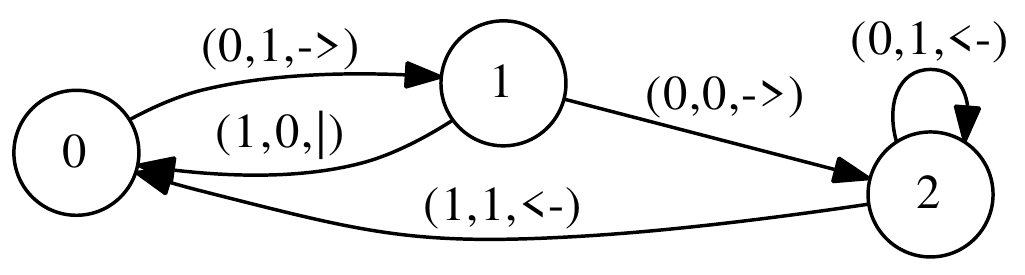}} %
\subfigure["$1^{6}$",13]{\includegraphics[scale=0.6]{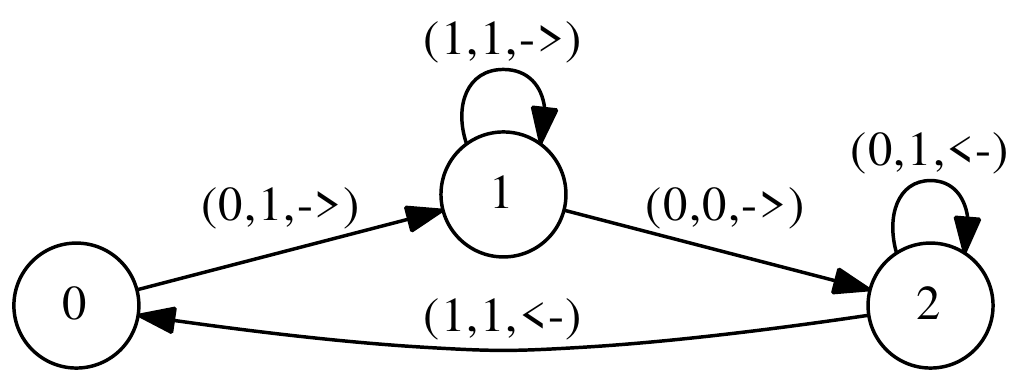}} \\ %
\subfigure["$1^{6}$",16]{\includegraphics[scale=0.6]{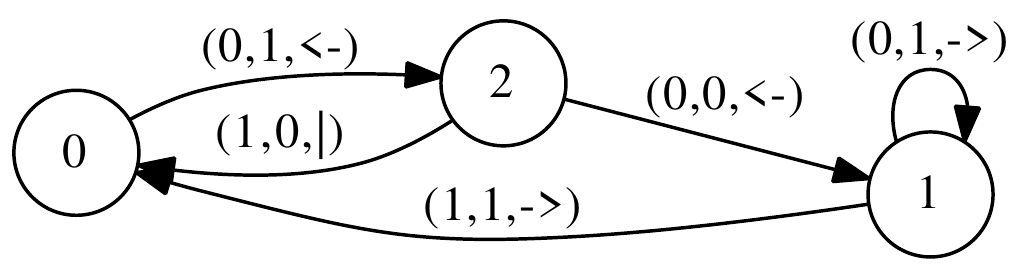}} %
\subfigure["$1^{6}$",13]{\includegraphics[scale=0.6]{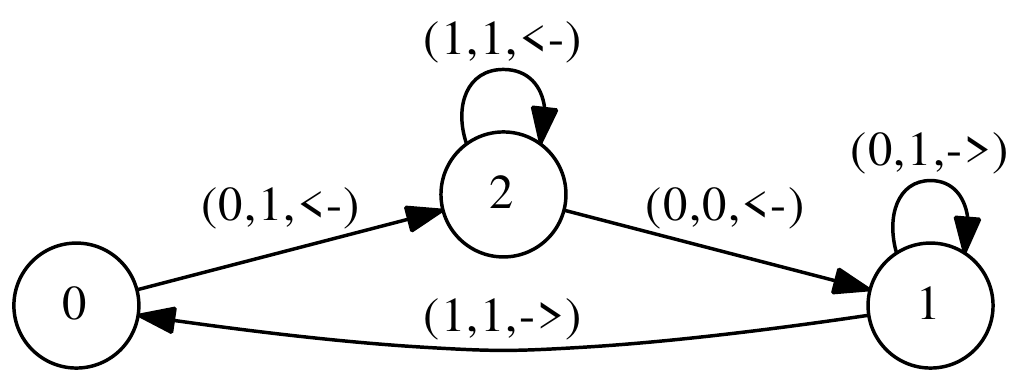}} 
\caption{All $\mathcal{M}_{BB}(3)$ Busy Beaver Machines (30-39), ("$T(\lambda)$", \# of steps)}
\end{figure}

\begin{figure}[h!]
\centering
\subfigure["$1^{6}$",15]{\includegraphics[scale=0.6]{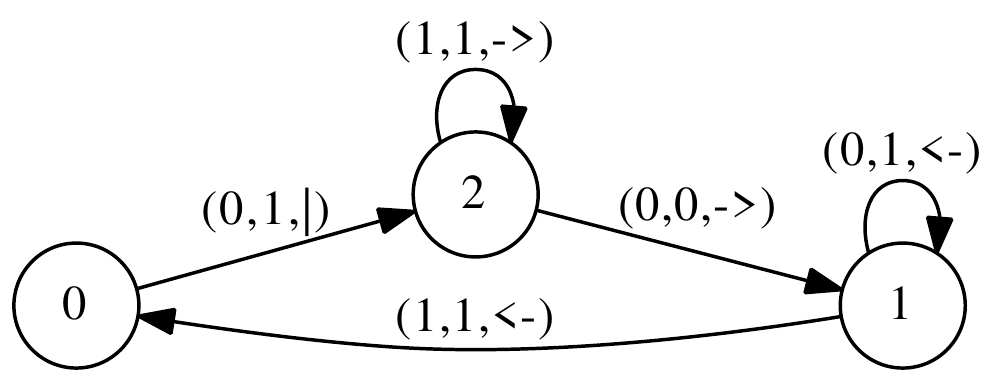}} %
\subfigure["$1^{6}$",15]{\includegraphics[scale=0.6]{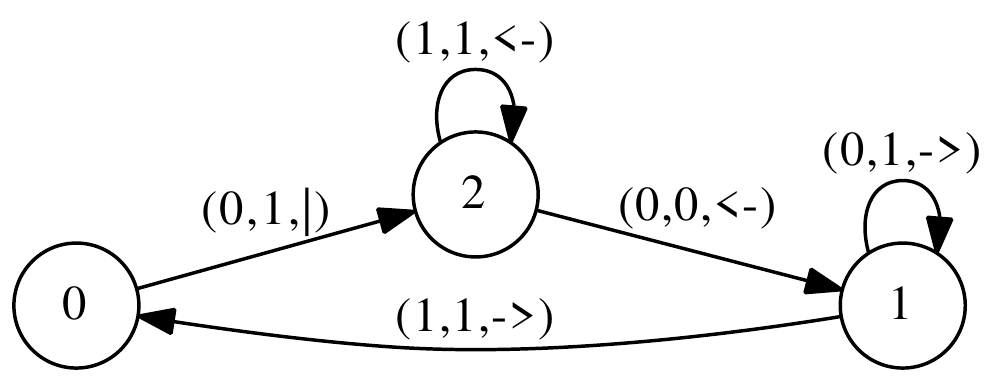}} \\ %
\subfigure["$1^{6}$",16]{\includegraphics[scale=0.6]{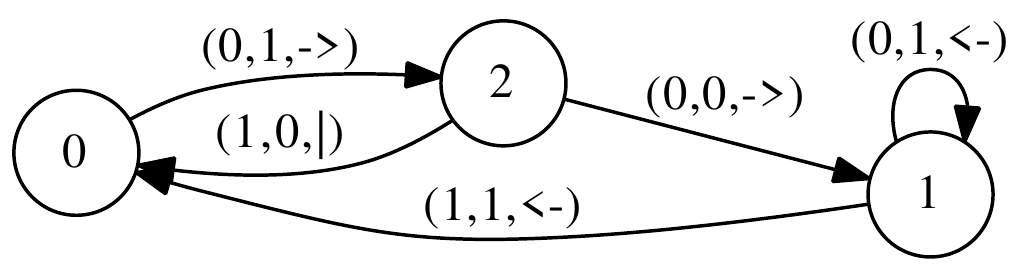}} %
\subfigure["$1^{6}$",13]{\includegraphics[scale=0.6]{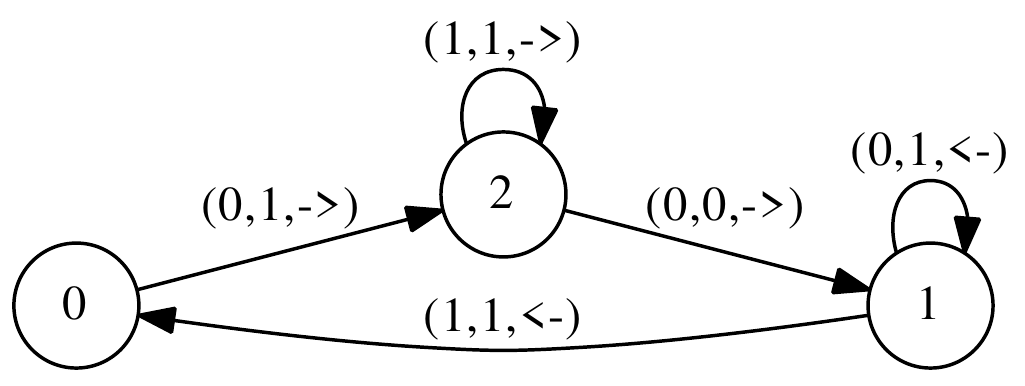}} 
\caption{All $\mathcal{M}_{BB}(3)$ Busy Beaver Machines (40-43), ("$T(\lambda)$", \# of steps)}
\end{figure}

\clearpage
\subsection{$\mathcal{M}_{BB}(4)$}
\label{mbb4}

\begin{figure}[h!]
\centering
\subfigure["$1^{12}$",106]{\label{mbb4m1}\includegraphics[scale=0.43]{bb_4_24_0}} %
\subfigure["$1^{12}$",95]{\includegraphics[scale=0.43]{bb_4_24_1}} \\ %
\subfigure["$1^{12}$",106]{\label{mbb4m3}\includegraphics[scale=0.43]{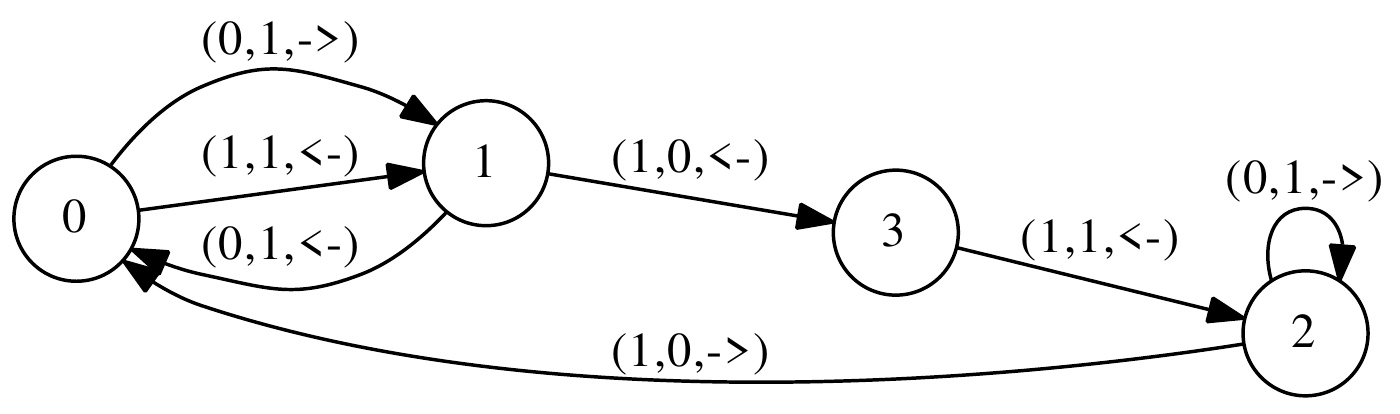}} %
\subfigure["$1^{12}$",95]{\includegraphics[scale=0.43]{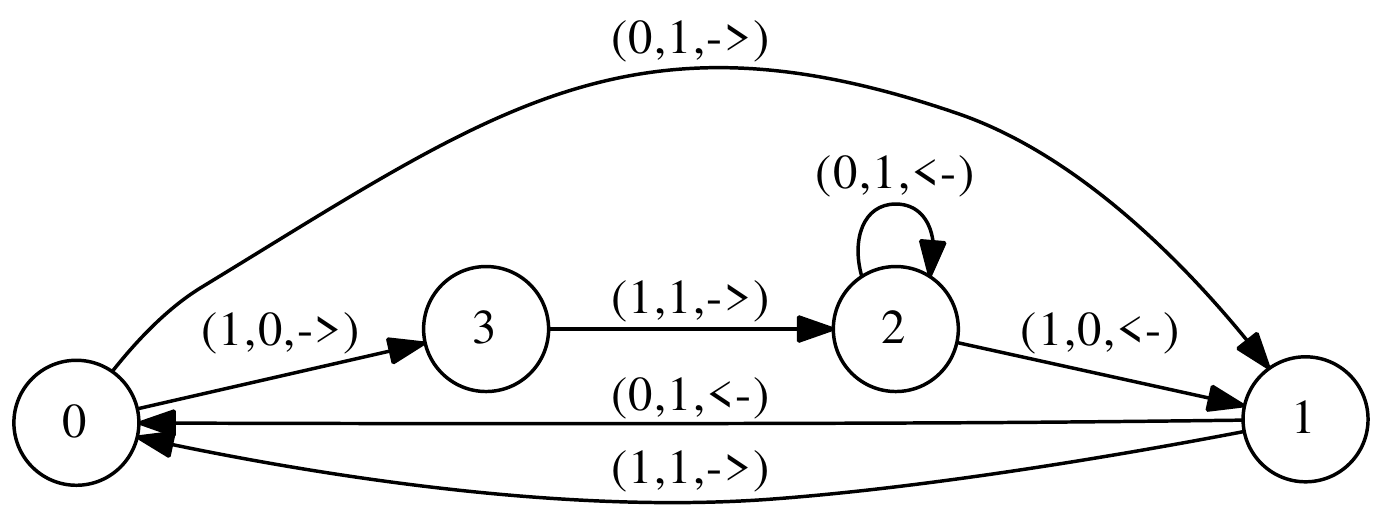}} \\ %
\subfigure["$1^{12}$",106]{\label{mbb4m5}\includegraphics[scale=0.43]{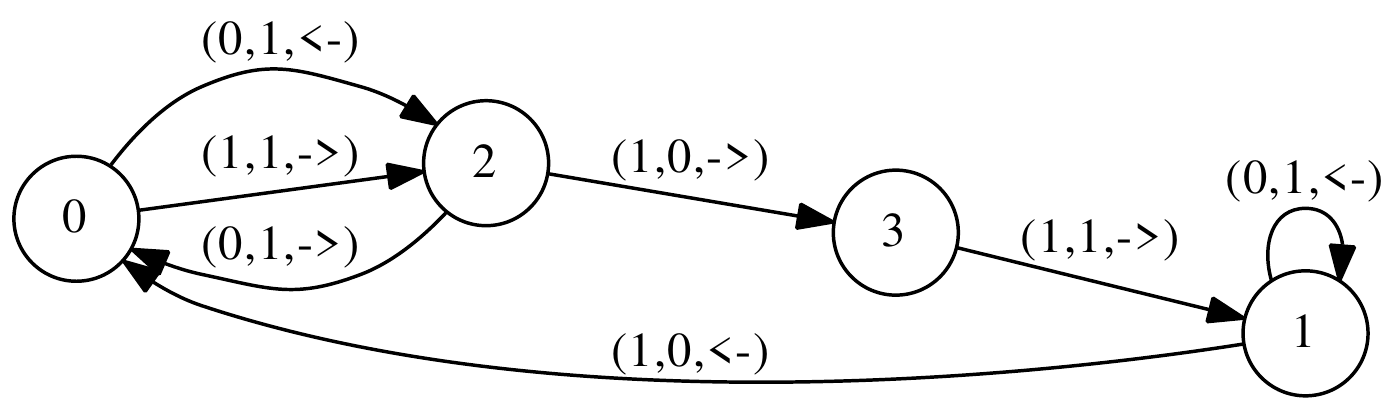}} %
\subfigure["$1^{12}$",95]{\includegraphics[scale=0.43]{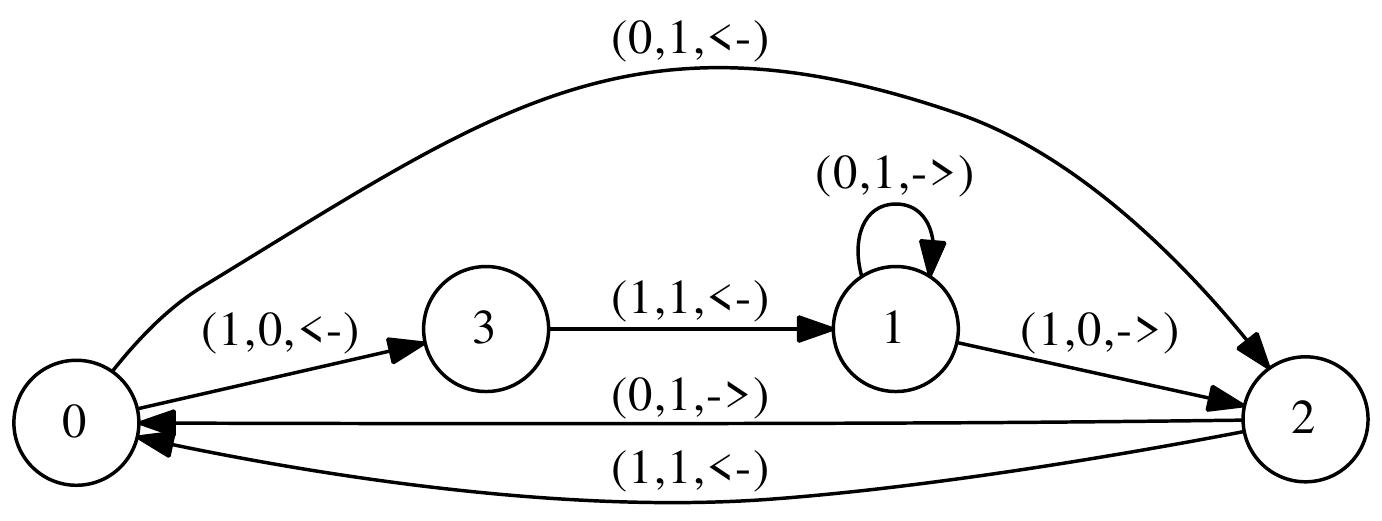}} \\ %
\subfigure["$1^{12}$",106]{\includegraphics[scale=0.43]{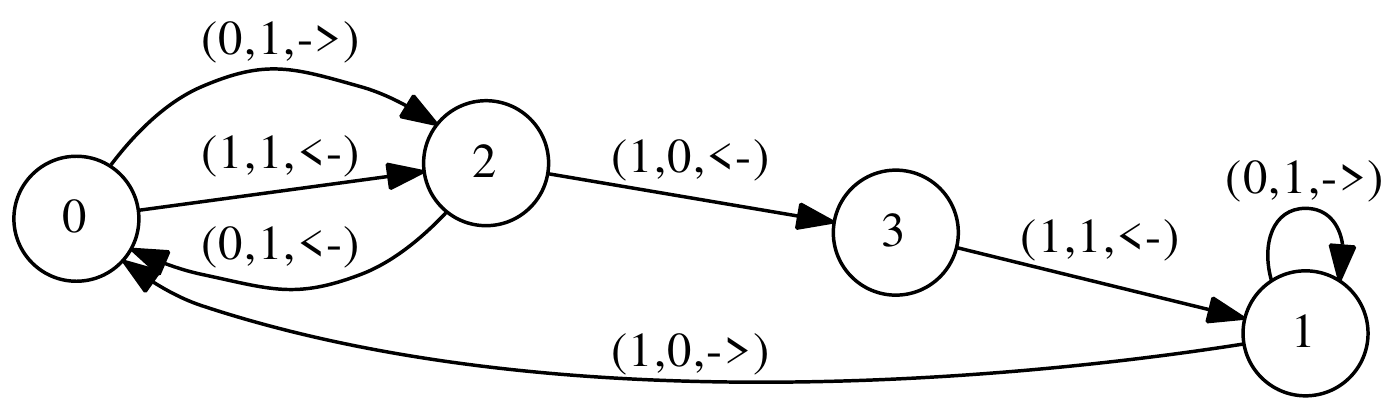}} %
\subfigure["$1^{12}$",95]{\includegraphics[scale=0.43]{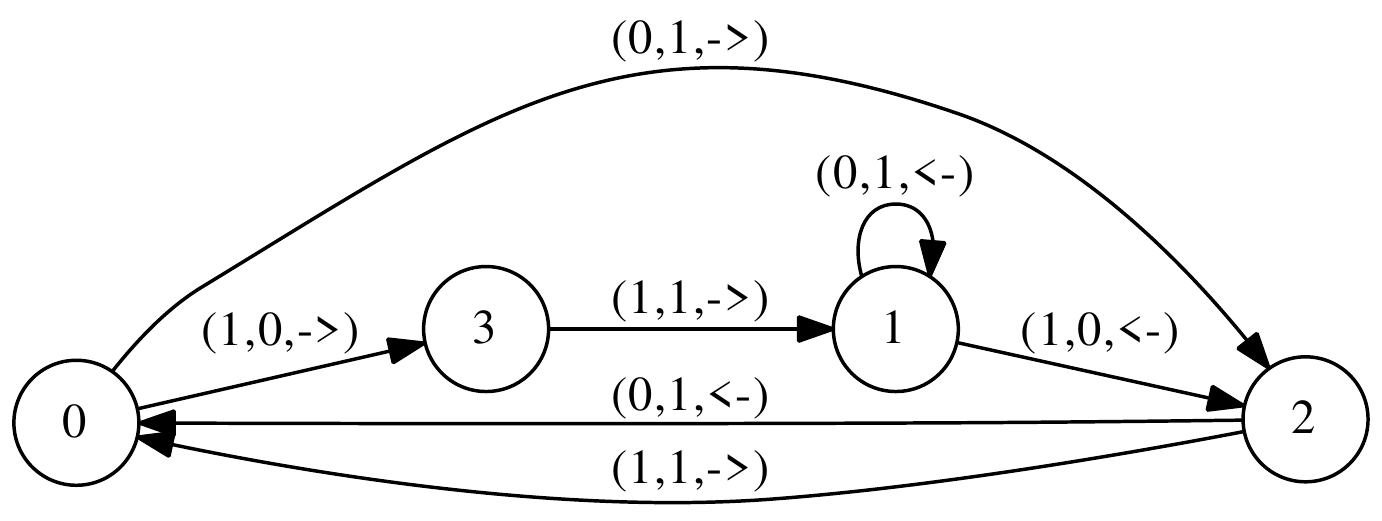}} \\ %
\subfigure["$1^{12}$",106]{\includegraphics[scale=0.43]{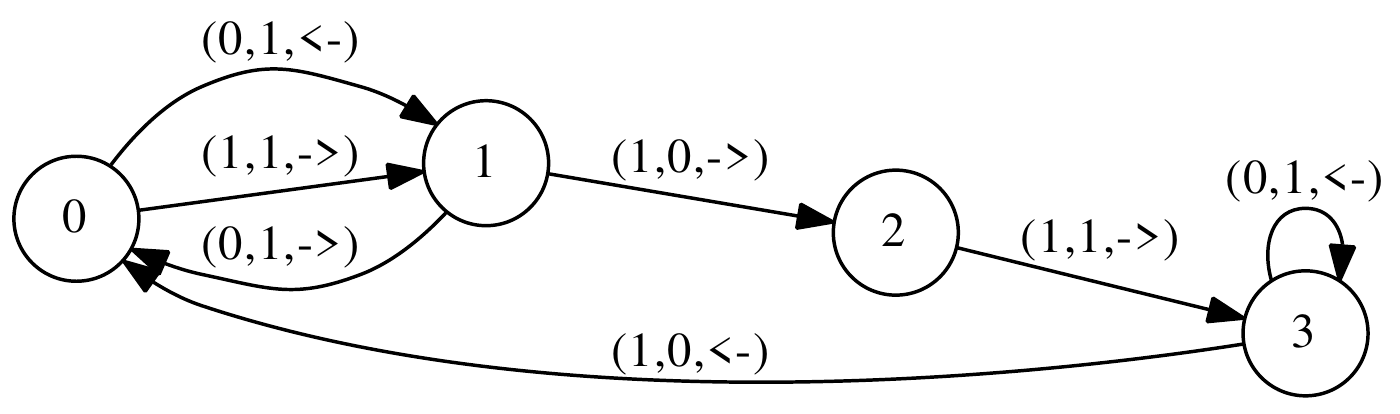}} %
\subfigure["$1^{12}$",95]{\includegraphics[scale=0.43]{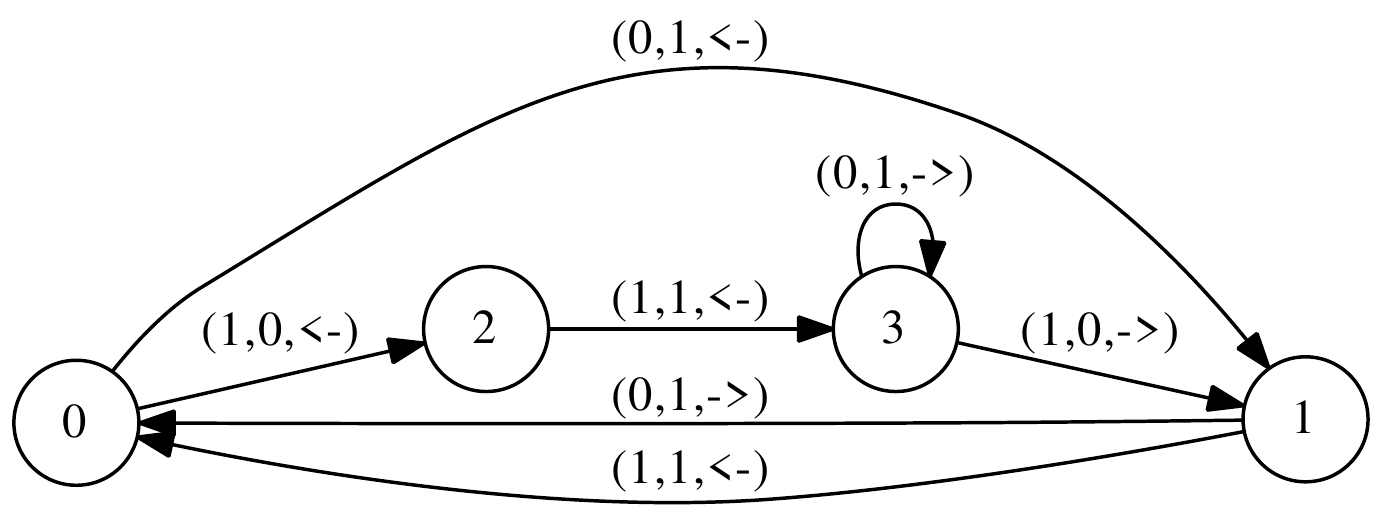}} 
\caption{All $\mathcal{M}_{BB}(4)$ Busy Beaver Machines (0-9), ("$T(\lambda)$", \# of steps)}
\end{figure}

\begin{figure}[h!]
\centering
\subfigure["$1^{12}$",106]{\includegraphics[scale=0.43]{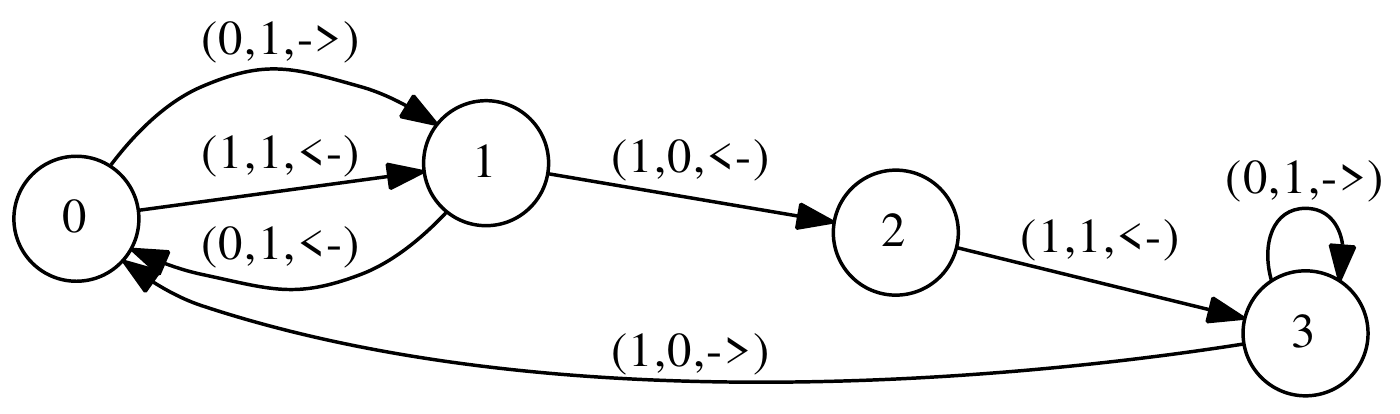}} %
\subfigure["$1^{12}$",95]{\includegraphics[scale=0.43]{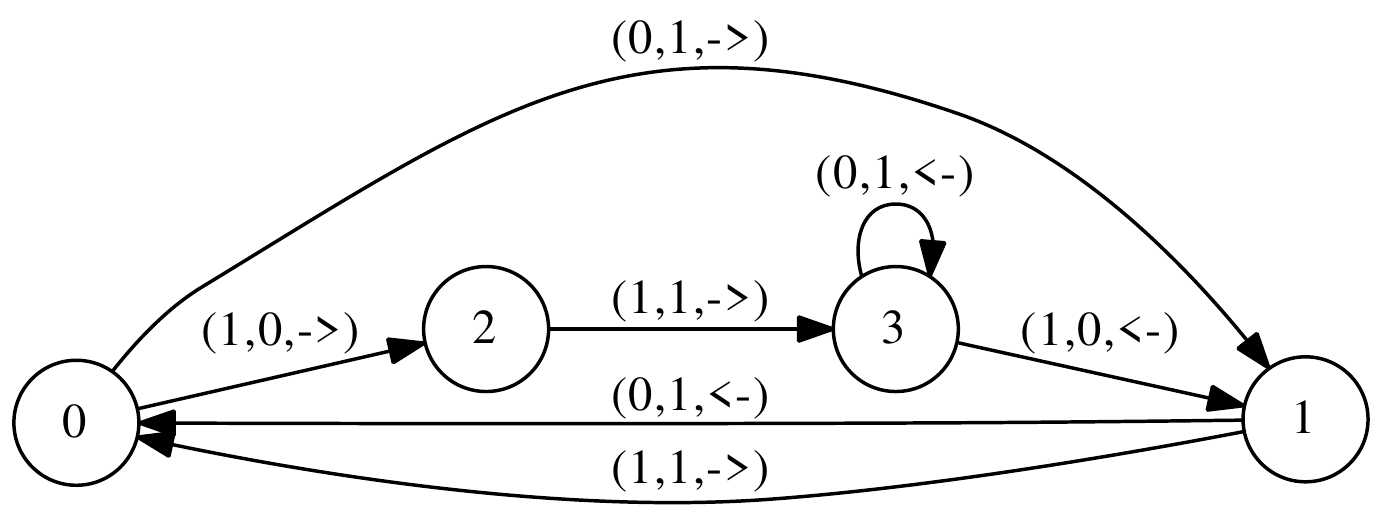}} \\ %
\subfigure["$1^{12}$",95]{\includegraphics[scale=0.43]{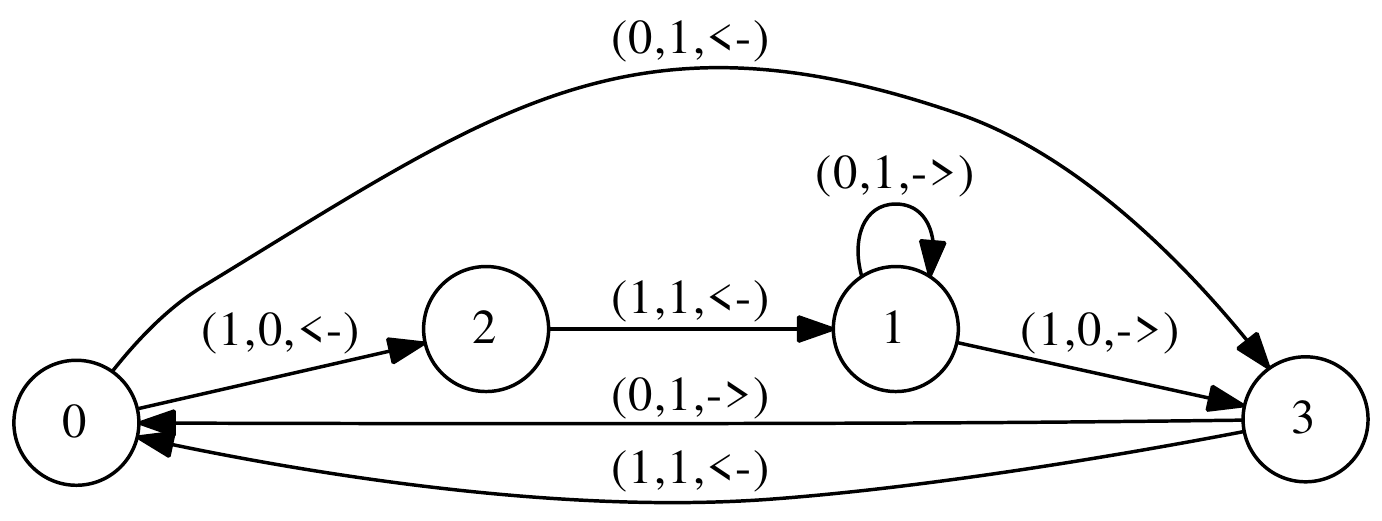}} %
\subfigure["$1^{12}$",106]{\includegraphics[scale=0.43]{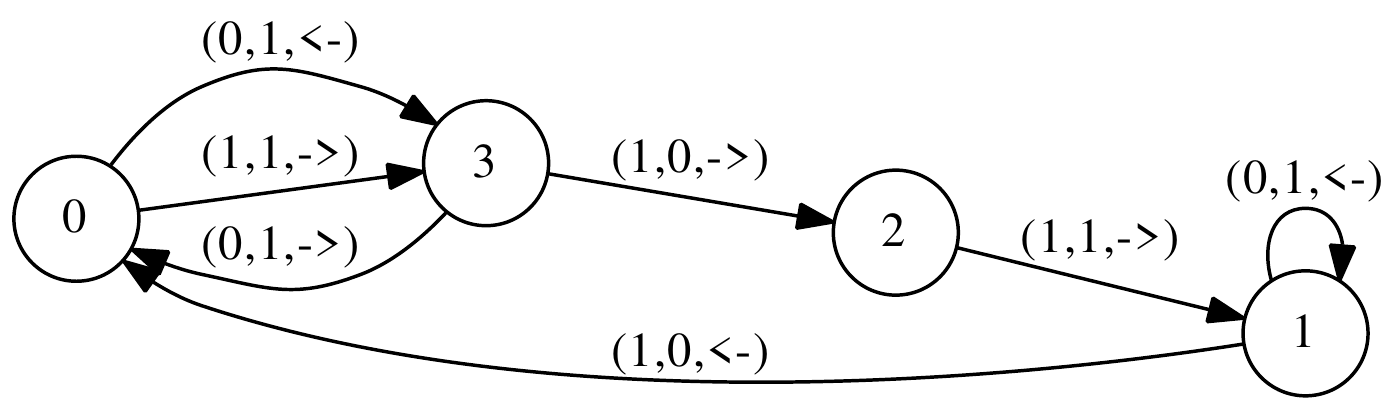}} \\ %
\subfigure["$1^{12}$",95]{\includegraphics[scale=0.43]{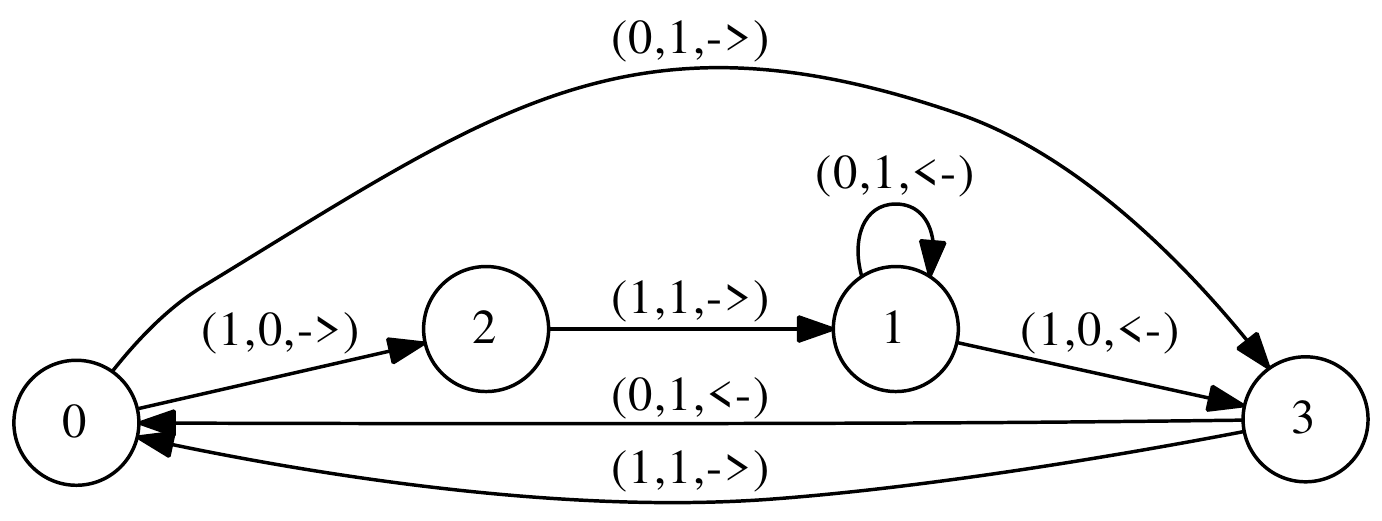}} %
\subfigure["$1^{12}$",106]{\includegraphics[scale=0.43]{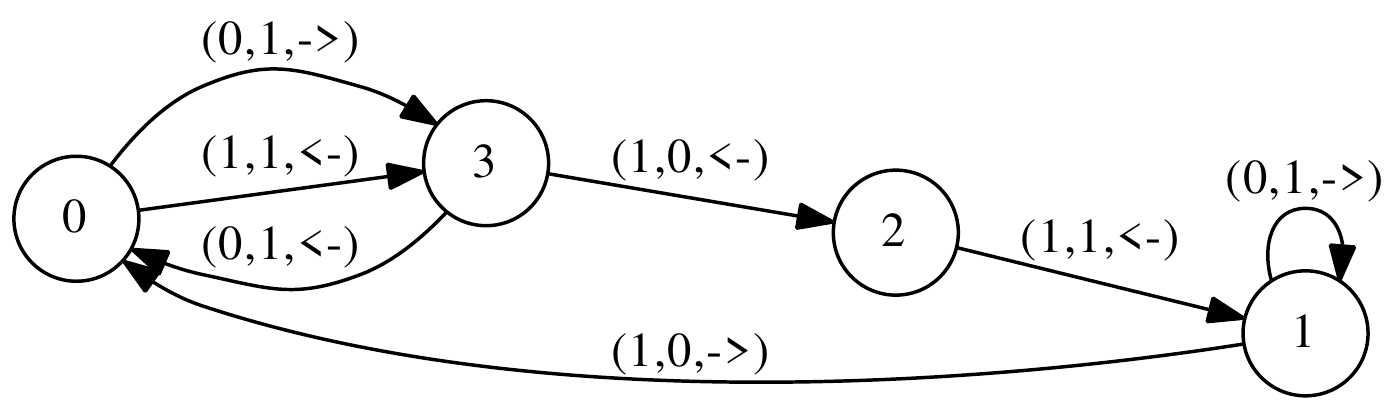}} \\ %
\subfigure["$1^{12}$",95]{\includegraphics[scale=0.43]{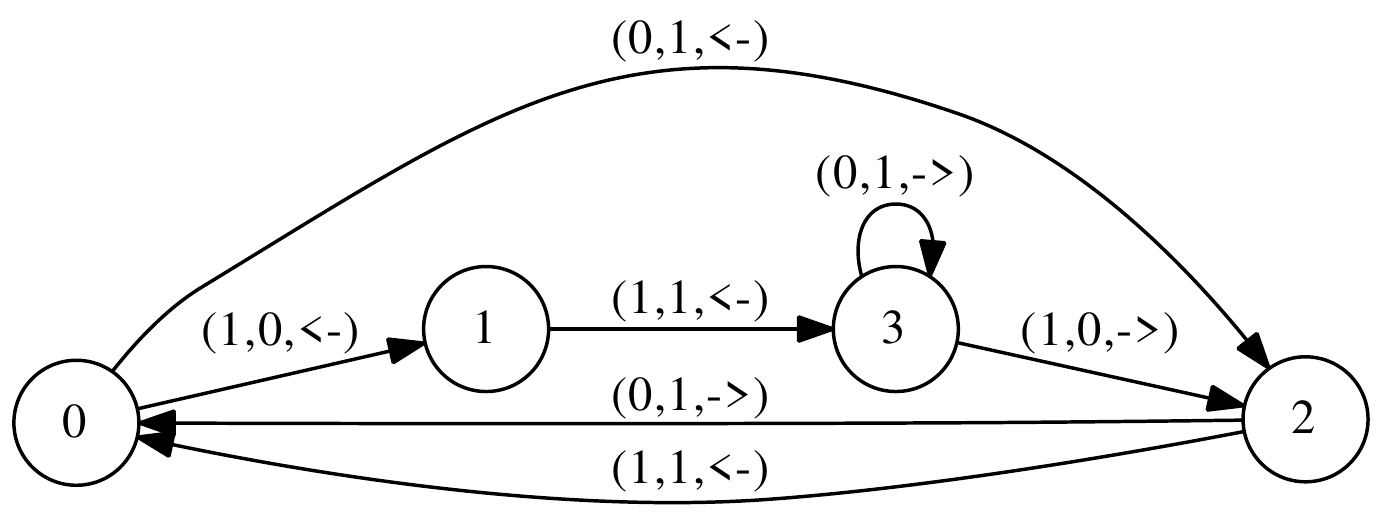}} %
\subfigure["$1^{12}$",106]{\includegraphics[scale=0.43]{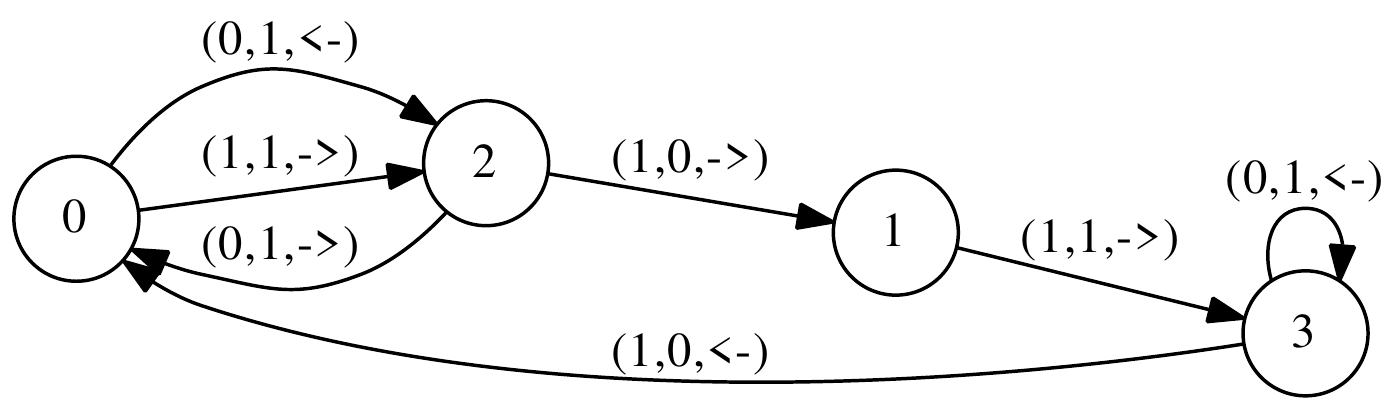}} \\ %
\subfigure["$1^{12}$",95]{\includegraphics[scale=0.43]{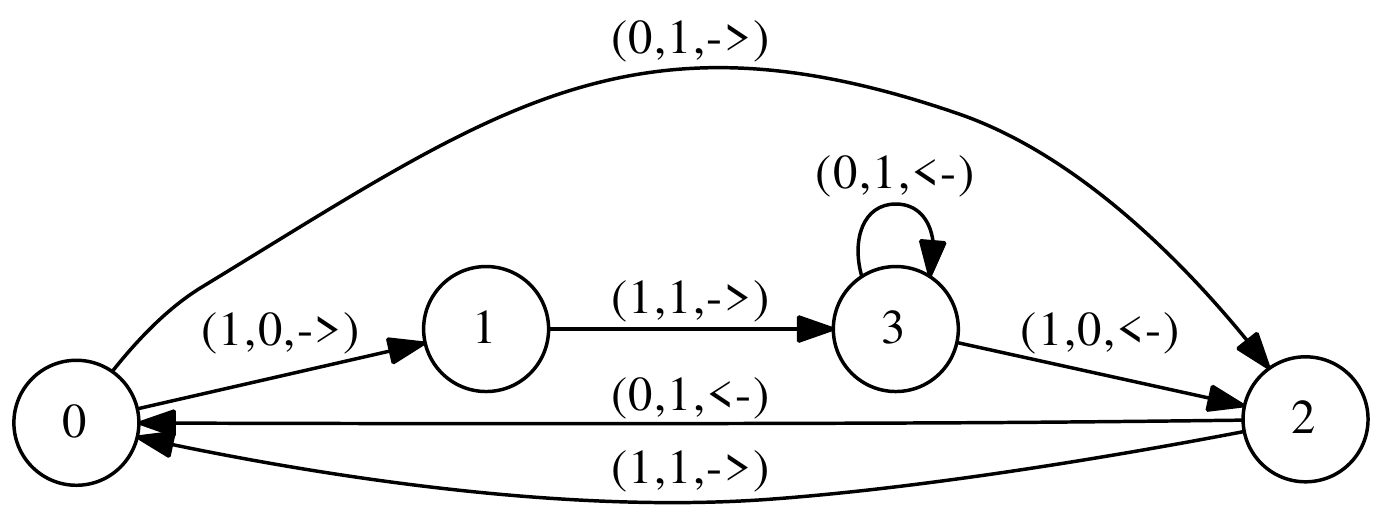}} %
\subfigure["$1^{12}$",106]{\includegraphics[scale=0.43]{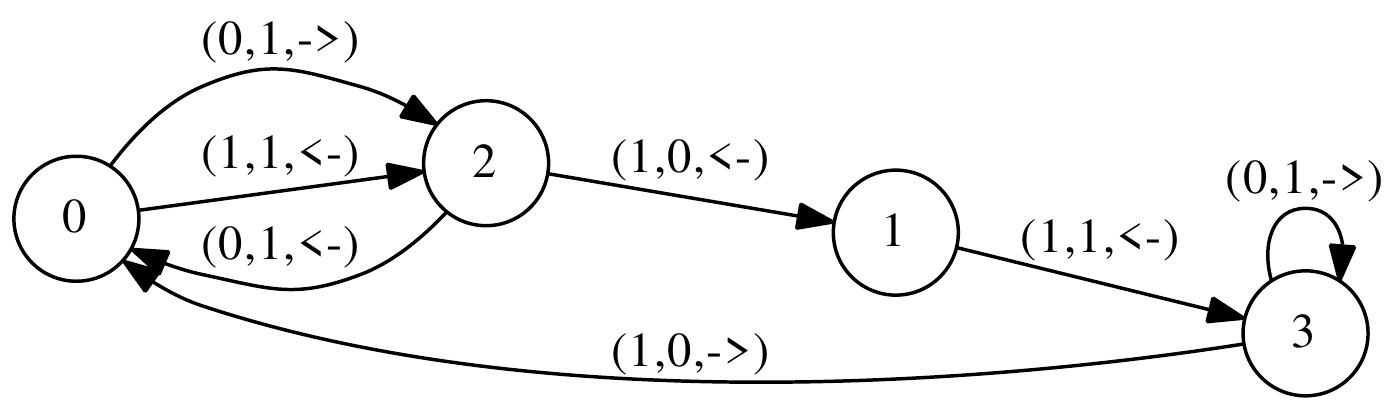}} 
\caption{All $\mathcal{M}_{BB}(4)$ Busy Beaver Machines (10-19), ("$T(\lambda)$", \# of steps)}
\end{figure}

\begin{figure}[h!]
\centering
\subfigure["$1^{12}$",95]{\includegraphics[scale=0.43]{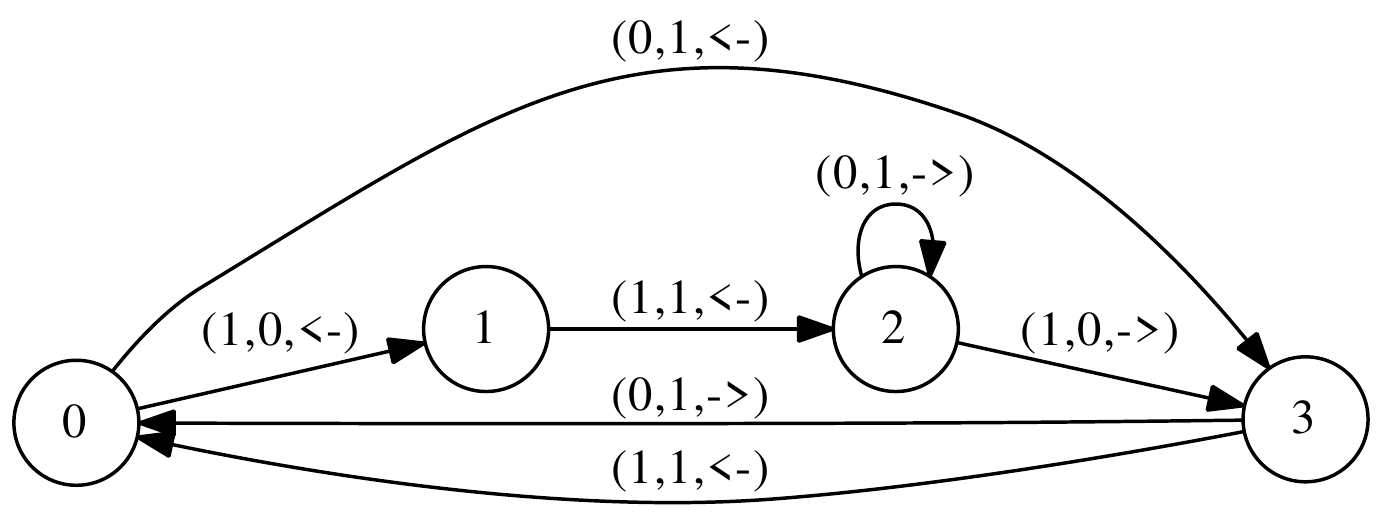}} %
\subfigure["$1^{12}$",106]{\includegraphics[scale=0.43]{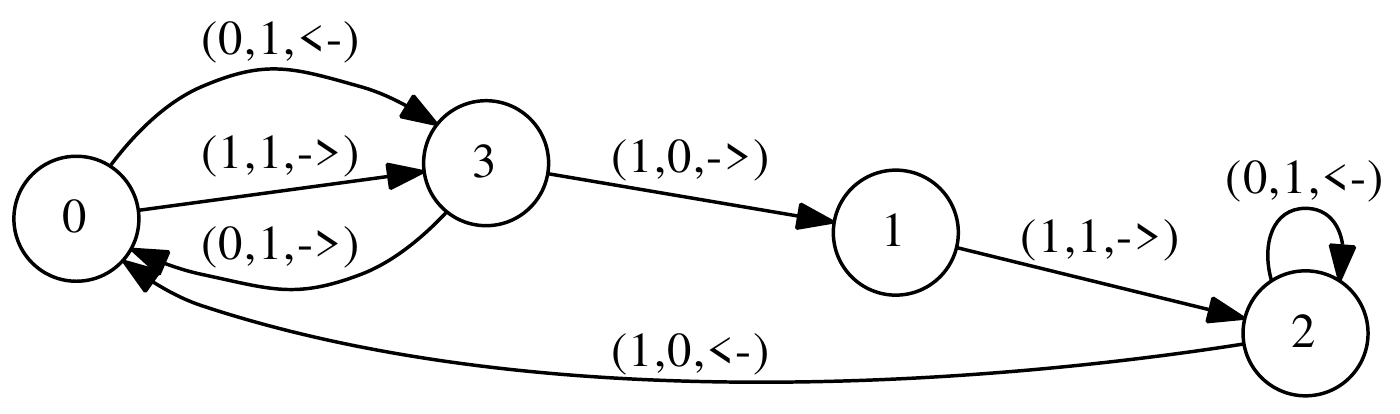}} \\ %
\subfigure["$1^{12}$",95]{\includegraphics[scale=0.43]{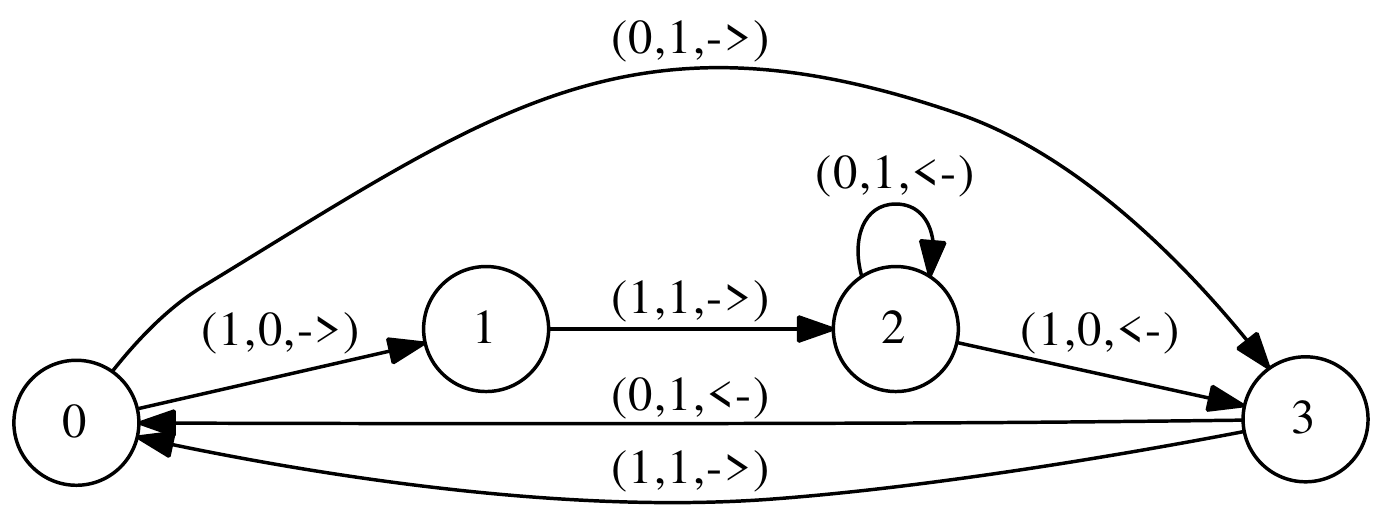}} %
\subfigure["$1^{12}$",106]{\includegraphics[scale=0.43]{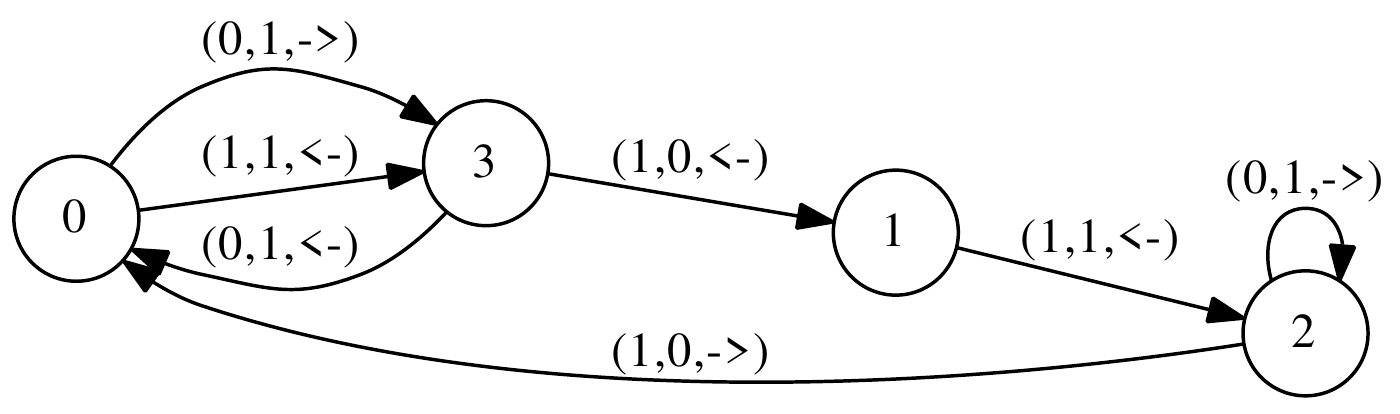}} \\ %
\caption{All $\mathcal{M}_{BB}(4)$ Busy Beaver Machines (20-23), ("$T(\lambda)$", \# of steps)}
\end{figure}

\clearpage
\section{Some C code snipets}

\subsection{$BB_5$}
\label{codesnips}

The array indexes contained in this code snipet may be used to decrypt the names of Turing machines. For example, from the name $(n, f_1, t_1, \dots,$ $2, 18,$ $\dots, f_n, t_n)$ with indexes $2$ and $18$ we can decrypt the rule $(1,0) \rightarrow (3,0,\leftarrow)$.

\begin{verbatim}
#define NOF_FROM 10
#define NOF_TO 30

/* (state, read) -> */
int from[NOF_FROM][2] = {
  {0, 0}, // 0
  {0, 1}, // 1
  {1, 0}, // 2
  {1, 1}, // 3
  {2, 0}, // 4
  {2, 1}, // 5
  {3, 0}, // 6
  {3, 1}, // 7
  {4, 0}, // 8
  {4, 1}  // 9
};

/* -> (state, write, move) 
                     0: <-
                     1: |
                     2: -> */
int to[NOF_TO][3] = {
  {0, 0, 0}, // 0
  {0, 0, 1}, // 1
  {0, 0, 2}, // 2
  {0, 1, 0}, // 3
  {0, 1, 1}, // 4
  {0, 1, 2}, // 5
  {1, 0, 0}, // 6
  {1, 0, 1}, // 7
  {1, 0, 2}, // 8
  {1, 1, 0}, // 9
  {1, 1, 1}, // 10
  {1, 1, 2}, // 11
  {2, 0, 0}, // 12
  {2, 0, 1}, // 13
  {2, 0, 2}, // 14
  {2, 1, 0}, // 15
  {2, 1, 1}, // 16
  {2, 1, 2}, // 17
  {3, 0, 0}, // 18
  {3, 0, 1}, // 19
  {3, 0, 2}, // 20
  {3, 1, 0}, // 21
  {3, 1, 1}, // 22
  {3, 1, 2}, // 23
  {4, 0, 0}, // 24
  {4, 0, 1}, // 25
  {4, 0, 2}, // 26
  {4, 1, 0}, // 27
  {4, 1, 1}, // 28
  {4, 1, 2}  // 29
};

...

     if ((nr_1s = 
          simulate (9, f1, t1, f2, t2, f3, t3, f4, 
                    t4, f5, t5, f6, t6, f7, t7, f8,
                    t8, f9, t9)) > max_ones)
       {...
...
\end{verbatim}

\clearpage
\section{Acknowledgement}
The author would like to thank Professor Gy\"orgy Terdik for encouraging starting of research in algorithmic information theory in our department. Since Spring 2009, we together organize a learning seminar on Kolmogorov Complexity. We are interested in developing programs which are based on data and CPU intensive computing. 

\bibliography{OnTheRunning}

\end{document}